\documentclass[11pt]{article}
\begin{document}

\title {ONE POSSIBLE INTERACTION-INERTIAL INTERACTION}
\author{Yang Xuejun\thanks{E-mail:yang\_xue\_jun@163.com },
\\
\centerline{\textit{Department of Physics, Shaoxing University, }}
\\
\centerline{\textit{Shaoxing 312000, P. R. China}}
\\
\\
} \maketitle

\begin{abstract}
Proposed in this paper is a possible interaction which exists in
nature - inertial interaction. It gives matter an inertia and
inertial mass. The formula of inertial mass has been derived. It
is possible that inertial interaction leads to the redshifts of
quasars, the rotation curve of spiral galaxy, the accelerating
expansion of the universe, and the stronger gravitational lens
effects of quasars, galaxies, or clusters of galaxies. Einstein's
Gravitational Equation has been modified. Gravitational redshift,
perihelion precession, and bending of light in spherically
symmetric vacuum gravitational field are calculated. The
differential equations of static spherically symmetric star's
internal evolution are given. The accelerating expansion stage of
the universe evolution equations are derived. The evolution of the
universe is periodic. Time does not have an origin. There is no
Big Bang. Although there is divergent singularity, there is no
universe's singularity of incomplete geodesic. There are no
horizon problem and no flatness problem. The problems that may
exist are discussed.
\end{abstract}
\section{INTRODUCTION}

It is predicted that perhaps there exists inertial interaction in
nature and the inertial interaction gives matter an inertia and
inertial mass. It is possible that inertial interaction leads to
the redshifts of quasars, the rotation curve of spiral galaxy, the
accelerating expansion of the universe, and the stronger
gravitational lens effects of quasars, galaxies, or clusters of
galaxies.

In Newton's bucket experiment\cite{zhaozheng}, how can the water
know that only when it rotates relative to the distant galaxies in
the universe, instead of relative to the bucket wall, the
concavity of the water surface changes from convex to concave? It
can be assumed that the concave water surface is the result of the
interaction of distant galaxies in the universe to the water; or,
the inertia of matter is the result of the interaction of other
substances to the matter. Let us imagine that the entire
universe's substance is electrically neutral and distributed as a
uniform spherical shell. There is an object in the universe's
spherical shell. The inertia of the object is the result of the
interaction of spherical shell universe's substance to the object
itself. The interaction of the object is obviously not from the
four fundamental interactions of the spherical shell universe. A
possible alternative is that, let us imagine, this kind of
interaction is a new interaction which is different from any one
of the four fundamental interactions. Or, at least, it is another
aspect of the gravitational interaction which people have not
recognized. If such an interaction exists, we can call it inertial
interaction or gravitational inertial interaction.

\section{THE FORMULA OF INERTIAL MASS}

\subsection{THE CONTRIBUTION OF INERTIAL MASS FROM COSMIC BACKGROUND}

Inertial interaction gives matter an inertia and inertial mass.
Let $M_{I12}$ and $M_{I21}$ be respectively the inertial masses of
particle 1 to particle 2 and particle 2 to particle 1. Assume
\begin{equation}
M_I\equiv M_{I12}=M_{I21}=K\frac{M_{G1}M_{G2}}{r^n}exp(-\delta r).
\label{gxzlgs1}
\end{equation}
where, $M_{G1}$ and $M_{G2}$ are the gravitational masses of
particle 1 and particle 2 respectively. $r$ is the length of the
geodesic between particle 1 and particle 2. $K(>0)$ and $\delta
(>0)$ are the constants to be determined. $n$ is an integer to be
determined. If the inertial interaction is weaker than the
gravitational interaction and if it is long-range, $K$ and $\delta
$ will be very small.

The following calculation is for the inertial mass of the
particle, which is the result of the inertial interaction of the
cosmic background to the particle with gravitational mass $M_G$.
Let us assume that the cosmological principle holds, then the
universe metric is Robertson-Walker metric\cite{liangchanbin}.
\begin{equation}
ds^2=-dt^2+a^2(t)[\frac{dr^2}{1-kr^2}+r^2(d\theta^2+sin^2\theta
d\varphi^2)]. \label{R-W}
\end{equation}
Let us consider a flat space($k=0$) (Other cases can also be
calculated in the same manner.). The distance between the two
points $A(r_A,\theta,\varphi)$ and $B(r_B,\theta,\varphi)$
is\cite{liangchanbin}
\begin{equation}
D_{AB}(t)=a(t)\int^{r_B}_{r_A}dr/\sqrt{1-kr^2}=a(t)(r_B-r_A).
\label{juli}
\end{equation}
Let the origin point of Robertson-Walker coordinate be the point
where the particle sits, and the volume element of the point
$(r,\theta,\varphi)$ is\cite{liangchanbin}
\begin{equation}
\hat{\varepsilon}=a^3(t)r^2sin\theta dr\Lambda d\theta\Lambda
d\varphi. \label{tiyuan}
\end{equation}
The inertial mass of the particle, which is the result of the
inertial interaction of the cosmic background to the particle, is
\begin{equation}
\begin{array}{l}
M'_I(t)=\int_{\Sigma}K\frac{M_G}{(ar)^n}exp(-\delta ar)\rho_G \hat{\varepsilon}\\
=K\frac{4\pi M_G \rho_G}{\delta^{3-n}}\Gamma(3-n)\equiv \alpha_U M_I\\
\end{array}
\label{yzbjgxzyzl}
\end{equation}
where, $\Gamma(3-n)$ is $\Gamma$ function. $\alpha_U$ is the
contribution rate of cosmic background for inertial mass of the
particle. $\Sigma$ is the space of universe at moment $t$.
$\rho_G$ is the gravitational mass density of the universe at
moment $t$. $\alpha_U$ is the function of time $t$ through
$\rho_G$. On the Earth today $M_I=M_G$, From equation
(\ref{yzbjgxzyzl}), we can obtain

\begin{equation}
K\frac{4\pi \rho_G}{\alpha_U\delta^{3-n}}\Gamma(3-n)=1.
\label{xishu}
\end{equation}
The inertial mass of the particle on the Earth is mainly
contributed by the cosmic background, the Milky Way, the Sun and
the Earth's inertial interaction. Let $\alpha_U$, $\alpha_M$,
$\alpha_\odot$, and $\alpha_E$ be the contribution rates
respectively. Today $\alpha_U+\alpha_M+\alpha_\odot+\alpha_E=1$.

\subsection{THE ROTATION CURVE OF SPIRAL GALAXY AND THE FORMULA OF INERTIAL MASS}

The inertial mass of a star in a spiral galaxy is mainly
contributed by the background of the universe and the spiral
galaxy through inertial interaction. The inertial mass of the star
is
\begin{equation}
M_{IS}=\alpha_U M_G+K\frac{M_{GSg}M_G}{(r_S)^n}exp(-\delta r_S).
\label{hxgxzl}
\end{equation}
where, $M_{GSg}$ is the gravitational mass of the luminous part of
the spiral galaxy. $r_S$ is the distance from the star to the
center of the spiral galaxy. $M_G$ is the gravitational mass of
the star. If $n=1$, $\delta r_S\ll 1$, and $\alpha_U\ll
K\frac{M_{GSg}}{r_S}$, at the outside of the core of the spiral
galaxy, from $M_{IS}\frac{V_0^2}{r}=G\frac{M_{GSg}M_G}{r^2}$ and
equation (\ref{hxgxzl})) we can obtain (Here, equation
(\ref{ndlxfc1}) is used)
\begin{equation}
V_0=\sqrt{G\frac{M_{GSg}M_G}{rM_{IS}}}\simeq\sqrt{\frac{G}{K}}.
\label{hwxzsd}
\end{equation}
where, $V_0$ is the velocity of the star rotating around the core
of the spiral galaxy. From equation (\ref{hwxzsd}) we know that
$V_0$ is approximately a constant and $V_0$ is independent of the
gravitational mass of the light-emitting part of the spiral
galaxy. Almost all the $V_0$ of all spiral galaxies are the same.
The formula (\ref{gxzlgs1}) of inertial mass between particle 1
and particle 2 becomes
 \begin{equation}
M_I\equiv M_{I12}=M_{I21}=K\frac{M_{G1}M_{G2}}{r}exp(-\delta r).
\label{gxzlgs2}
\end{equation}
where $K$ can be obtained from equation (\ref{hwxzsd})
\begin{equation}
K=\frac{G}{V_0^2}. \label{K}
\end{equation}
Let $G=6.67\times 10^{-11}Nm^2/kg^2$ and $V_0=2\times 10^5m/s$,
then $K=1.67\times 10^{-21}m/kg$.

From calculation, we know that the inertial mass of a star at the
center of spiral galaxy is
$M_{IS}=\frac{3}{2}K\frac{M_{GSg}M_G}{r_0}$. We can also know that
the inertial mass of the star at the edge of spiral galaxy core is
$M_{IS}\simeq\frac{3}{4}K\frac{M_{GSg}M_G}{r_0}$, where $r_0$ is
the radius of the spiral galaxy core. Therefore, it can be
approximately regarded that the inertial mass of a star in the
spiral galaxy nucleus has nothing to do with the location within
spiral galaxy nucleus of the star. From
$M_{IS}\frac{V_0^2}{r}=G\frac{M_{GSr}M_G}{r^2}$, we have the speed
of rotation of a star in the spiral galaxy nucleus
\begin{equation}
V_0=\sqrt{G\frac{M_{GSr}M_G}{rM_{IS}}}\simeq\sqrt{G\frac{4\pi\rho_GM_G}{3M_{IS}}}r\propto
r. \label{hnxzsd}
\end{equation}
where, $M_{GSr}$ is the gravitational mass of the spiral galaxy
nucleus in the range of radius $r$. $\rho_G$ is the gravitational
mass density of the spiral galaxy nucleus.

In addition, $\frac{1}{M_I}\frac{dM_I}{dr}= -\frac{1}{r}-\delta$
can be obtained by using the formula (\ref{gxzlgs2}) of inertial
mass. On the scale of galaxy clusters, the inertial mass $M_I'$ of
the gas molecules within the cluster of galaxies almost is
$M_I'\simeq\alpha_U M_I\approx 0.09M_I$. $M_I$ is the inertial
mass of the same molecule on the Earth. The value of $\alpha_U$
can be found in $\S 2.3$ at the latter part. On one hand, by the
following equation (\ref{guanxinpaichili}) we know that as a
result of the gravitation the escape deceleration $ \frac{d{\bf
V}}{dt}=\frac{1}{M_I'}{\bf F}- \frac{1}{M_I'}{\bf
V}\frac{dM_I'}{dt} \approx \frac{1}{\alpha_U M_I}{\bf F}
\sim\frac{1}{0.09}\frac{1}{M_I}{\bf F}$ of the hot gases in galaxy
clusters should be larger than we had expected. On the other hand,
$\frac{1}{2}\times0.09M_I\overline{v^2}\sim\frac{3}{2}K_BT$ and
$K_B$ is the Boltzmann constant. The thermal motion velocity
$\sqrt{\overline{v^2}} \sim
\frac{1}{\sqrt{0.09}}\sqrt{\frac{3}{M_I}K_BT}$ of the molecular is
also larger than we had expected. However, it is not larger than
the escape deceleration which is larger than we had expected.

\subsection{QUASAR REDSHIFT}

The inertial mass of a particle on the surface of a star in or
near the Milky Way is mainly contributed by the cosmic background,
the Milky Way, and the star through the inertia interaction. The
inertial mass of the particle is
\begin{equation}
M'_I=\alpha_U M_G+K\frac{M_{GM}M_G}{r_M}+K\frac{M_{GS}M_G}{R_S}
=(\alpha_U+K\frac{M_{GM}}{r_M}+K\frac{M_{GS}}{R_S})M_I.
\label{hxbmzdgxzl}
\end{equation}
where, $M_{GM}$ is the gravitational mass of the Milky Way, $r_M$
is the distance from the star to the galactic center, $M_{GS}$ is
the gravitational mass of the star, $R_S$ is the radius of the
star, $M_G$ is the gravitational mass of the particle, $M_I(=M_G)$
is the inertial mass of the particle on the Earth. If
$\alpha_U+K\frac{M_{GM}}{r_M}+K\frac{M_{GS}}{R_S}<1$, the electron
inertial mass on the surface of a star is less than the electron
inertial mass on the Earth and the Rydberg constant on the surface
of a star will be less than the Rydberg constant on the Earth, and
then this will lead to the redshift of Stellar spectrum. For a
star that is far from the galactic center, if
$K\frac{M_{GS}}{R_S}$ is not big enough the spectrum possibly is
redshift. If $K\frac{M_{GS}}{R_S}$ is big enough, resulting in
$\alpha_U+K\frac{M_{GM}}{r_M}+K\frac{M_{GS}}{R_S}>1$, then the
spectrum will possibly be violet-shift. There are possible
$\alpha_U+K\frac{M_{GM}}{r_M}+K\frac{M_{GS}}{R_S}<1$ and larger
redshifts for the stars, which are on the edge of the galaxy or
beyond the Milky Way, and the stars are possible quasars. If
$K\frac{M_{GM}}{r_M}\ll \alpha_U$ and $K\frac{M_{GS}}{R_S}\ll
\alpha_U$ so that $M'_I\simeq\alpha_UM_I$, then the quasar
redshift reaches maximum. Therefore, at least some of quasars are
the stars that are on the edge or away from the Milky Way. Due to
$M_{Iqe}\simeq(\alpha_U+\alpha_\odot)M_{Ie}$, the electron
inertial mass on the surface of the quasar that has the maximum
redshift ($M_{Ie}$ is the electron inertial mass on the Earth)
(here, as an approximation, $\alpha_\odot$ replaced
$K\frac{M_{GS}}{R_S}$), the Rydberg constant on the surface of the
quasar which has the maximum redshift is
$R_q=(\alpha_U+\alpha_\odot)R$ ($R$ is the Rydberg constant on the
Earth). The wavelength of the spectrum emitted by the element on
the surface of the quasar which has the maximum redshift is
$\lambda_q=\frac{1}{(\alpha_U+\alpha_\odot)}\lambda$ ($\lambda$ is
the wavelength of the spectrum of the same elements on the Earth).
The quasar maximum redshift is
$Z_{qmax}\equiv\frac{\lambda_q-\lambda}{\lambda}=\frac{1}{(\alpha_U+\alpha_\odot)}-1$,
then $\alpha_U+\alpha_\odot=\frac{1}{Z_{qmax}+1}$. Let the quasar
maximum redshift be $Z_{qmax}=5$, then
$\alpha_U+\alpha_\odot=\frac{1}{6}\approx 0.17$ and
$\alpha_U<0.17$. On the Earth
$\alpha_M\approx10\alpha_{\odot}\approx100\alpha_E$, then
$\alpha_M\approx 0.82$ can be obtained from
$\alpha_U+\alpha_M+\alpha_\odot+\alpha_E=1$, and current
$\alpha_U\approx 0.09$.

If a quasar is a distant active galaxy, all spectra of the quasar
have cosmological redshift. If quasars are the stars that are on
the edge of or away from the Milky Way, then only the emission
lines and absorption lines of the quasar by the electronic
transition in atom have redshifts and continuous spectrum have no
redshift. If the X-ray spectrum of a quasar is produced by
bremsstrahlung and the radiation intensity is inversely
proportional to the square of the inertial mass of the charged
particle, then the radiation intensity of a quasar's X-ray will be
approximately $10^2$ times as big as that of ordinary stars. If
the X-ray spectrum of a quasar is produced by synchrotron
radiation and the radiation intensity is inversely proportional to
$4$ power of the inertial mass of the charged particle, then the
radiation intensity of a quasar's X-ray will be approximately
$10^4$ times as big as that of ordinary stars.

From equation (\ref{yzbjgxzyzl}), we have $M'_I=K\frac{4\pi M_G
\rho_G}{\delta^2}\equiv \alpha_U M_I =\alpha_UM_G$. Therefore, the
cosmic gravitational mass density is
\begin{equation}
\rho_G(t)=\frac{\alpha_U \delta^2}{4\pi K}=\frac{\alpha_U \delta^2
V_0^2}{4\pi G}. \label{yzylzlmd}
\end{equation}
Due to $r_S\delta \ll 1$, let $\sigma r_S\delta\sim 1$, then
\begin{equation}
\rho_G(t)=\frac{\alpha_U V_0^2}{4\pi \sigma^2r_S^2G}.
\label{yzylzlmd1}
\end{equation}
Let $r_S\sim 10^{21}m$, $\alpha_U=0.09$, $\sigma=10^2$,and
$V_0=2\times 10^5m/s$, then the universe current gravitational
mass density is $\rho_G\sim 4.295\times 10^{-28}kg/m^3$, and the
universe current inertial mass density is
$\rho_I=\alpha_U\rho_G\sim 3.866\times 10^{-29}kg/m^3$.

\subsection{THE CYCLE OF A PENDULUM AT PERIHELION AND APHELION}

The formula (\ref{gxzlgs2}) of inertial mass can be verified by
measuring the cycle of a pendulum at Perihelion and aphelion. On
the Earth, the cycle of a pendulum is:
\begin{equation}
T=2\pi \sqrt{\frac{M_I}{M_G}\frac{l}{g}}. \label{danbaizhouqi}
\end{equation}
where, $M_I$ and $M_G$ are the inertial mass and gravitational
mass of the pendulum respectively. $l$ is the length of the swing
arm. $g$ is the Earth's gravitational acceleration. The inertial
mass of a body on the Earth is
$M_I=(\alpha_\odot+\alpha_U+\alpha_M+\alpha_E)M_G \simeq
(\frac{GM_\odot }{V_0^2r}+\alpha_U+\alpha_M++\alpha_E)M_G$, then
\begin{equation}
T=2\pi
\sqrt{(\frac{GM_\odot}{V_0^2r}+\alpha_U+\alpha_M++\alpha_E)\frac{l}{g}}.
\label{danbaizhouqi1}
\end{equation}
The cycle of a pendulum at Aphelion is
\begin{equation}
T_A=2\pi
\sqrt{(\frac{GM_\odot}{V_0^2(r_A+R_E)}+\alpha_U+\alpha_M++\alpha_E)\frac{l}{g_A}}.
\label{danbaizhouqiF}
\end{equation}
The cycle of a pendulum at Perihelion is
\begin{equation}
T_P=2\pi
\sqrt{(\frac{GM_\odot}{V_0^2(r_P-R_E)}+\alpha_U+\alpha_M++\alpha_E)\frac{l}{g_P}}.
\label{danbaizhouqiN}
\end{equation}
where, $r_A$ and $r_P$ are respectively the aphelion distance and
perihelion distance. $g_A$ and $g_P$ are the Earth's gravitational
acceleration at the aphelion and perihelion respectively. $R_E$ is
the radius of the Earth. From calculation, we have
\begin{equation}
\frac{T_P-T_A}{T_A}\sim 10^{-4}. \label{danbaizhouqixiangduicha}
\end{equation}

\section{THE MOTION EQUATION OF AN OBJECT}

\subsection{THE MOTION EQUATION OF AN OBJECT IN NEWTON'S SPACETIME}

First let us discuss the motion equation of an object in Newton's
spacetime. Let us assume that there exist only particle 1,
particle 2,..., particle $n$ and testing particle in whole space,
a total of the $n+1$ particles. Let ${\bf F}_1$, ${\bf
F}_2$...${\bf F}_n$ be respectively the forces applied on the
testing particle by the $n$ particles. Let $m_1$, $m_2$...$m_n$ be
respectively the inertial masses of the testing particle by the
$n$ particles through the inertial interaction. Let ${\bf r}_1$,
${\bf r}_2$...${\bf r}_n$ be respectively the position vectors of
the $n$ particles relative to the testing particle. Let ${\bf
V}_1\equiv \frac{d{\bf r}_1}{dt}$, ${\bf V}_2$...${\bf V}_n\equiv
\frac{d{\bf r}_n}{dt}$ be respectively the velocities of the $n$
particles relative to the testing particle. We assume that the
dynamical equations of the testing particle are the following
three possibilities:

1)
\begin{equation}
{\bf F}_1+{\bf F}_2+...+{\bf F}_n=-\frac{d}{dt}(m_1{\bf
V}_1+m_2{\bf V}_2+ ...+m_n{\bf V_n}). \label{ndlxfc}
\end{equation}
There are $n$ vectors ${\bf r}_1$, ${\bf r}_2$...,${\bf r}_n$ to
be determined. There are $n+1$ equations in the form of equation
(\ref{ndlxfc}) for the $n+1$ particles, but the sum of the $n+1$
equations is an identical equation $0=0$. Therefore, there are
only the $n$ independent equations in the $n+1$ equations. The $n$
vector equations are complete or self-sufficiency for the $n$
unknown vectors.

2)
\begin{equation}
{\bf F}_1+{\bf F}_2+...+{\bf F}_n=-\frac{d}{dt}(m_1{\bf
V}_1+m_2{\bf V}_2+ ...+m_n{\bf V}_n+m_{in}{\bf U}).
\label{ndlxfca}
\end{equation}
where, $m_{in}$ is the internal inertial interaction mass between
each internal component of the testing particle, and $m_{in}$ can
be called as the internal inertial mass of the testing particle.
${\bf U}$ is the velocity of the center of inertial masses of
particle 1, particle 2, ..., particle $n$ relative to the testing
particle.

3)
\begin{equation}
{\bf F}_1+{\bf F}_2+...+{\bf F}_n=-\frac{d(m{\bf U})}{dt}
=\frac{d(m{\bf V})}{dt}. \label{ndlxfcb}
\end{equation}
where, $m\equiv m_1+m_2+...+m_n+m_{in}$ is the total inertial mass
of the testing particle. ${\bf U}$ is the velocity of the center
of inertial masses of particle 1, particle 2, ..., particle $n$
relative to the testing particle. ${\bf V}\equiv -{\bf U}$ is the
velocity of the testing particle relative to the center of
inertial masses of particle 1, particle 2, ..., particle $n$.

Please note the following

[1] In equations (\ref{ndlxfc}) and (\ref{ndlxfca}), $\frac{d
(m_i{\bf V}_i)}{dt}$ is the inertial force applied to the testing
particle by the $i$-th particle. The testing particle also applies
an inertial force $-\frac{d (m_i{\bf V}_i)}{dt}$ to the $i$-th
particle. That is, inertial force satisfies Newton's third law.

[2] In the case of 1) and 2), we generally do not have the concept
of the inertial reference frame. But when ${\bf V}_1={\bf V}_2=
...={\bf V}_n\equiv -\bf V$, that is, no relative motions among
the $n$ particles, equations (\ref{ndlxfc}) and (\ref{ndlxfca})
become
\begin{equation}
{\bf F}_1+{\bf F}_2+...+{\bf
F}_n=\frac{d}{dt}((m_1+m_2+...+m_n){\bf V}) =\frac{d(m{\bf
V})}{dt}. \label{ndlxfc1}
\end{equation}
\begin{equation}
{\bf F}_1+{\bf F}_2+...+{\bf
F}_n=\frac{d}{dt}((m_1+m_2+...+m_n+m_{in}){\bf V}) =\frac{d(m{\bf
V})}{dt}. \label{ndlxfc1a}
\end{equation}
Referring to the above, $m\equiv m_1+m_2+...+m_n$ is the total
inertial mass of the testing particle as the result of the
inertial interaction of the $n$ particles. Or $m\equiv
m_1+m_2+...+m_n+m_{in}$ is the total inertial mass of the testing
particle. $\bf V$ is the velocity of the testing particle relative
to the $n$ particles. Then, equations (\ref{ndlxfc1}) and
(\ref{ndlxfc1a}) have the form of Newton's motion equation. In
this case, the system of the $n$ particles is the inertial
reference frame. Equation (\ref{ndlxfcb}) itself is a form of
Newton's motion equation and, in this case, the system of the $n$
particles is the inertial reference frame.

[3] In the case of 2) and 3), when $m_{in}$ cannot be ignored, the
star in spiral galaxy will deviate from the rotation curve of
spiral galaxy.

[4] If there only exist a testing particle and particle 1 in whole
space, equations (\ref{ndlxfc1}),(\ref{ndlxfc1a}), and
(\ref{ndlxfcb}) become
\begin{equation}
{\bf F}_1=\frac{d}{dt}(m_1{\bf V}). \label{ndlxfc2}
\end{equation}
\begin{equation}
{\bf F}_1=\frac{d}{dt}((m_1+m_{in}){\bf V}). \label{ndlxfc2a}
\end{equation}
\begin{equation}
{\bf F}_1=\frac{d}{dt}((m_1+m_{in}){\bf V}). \label{ndlxfc2b}
\end{equation}
Equations (\ref{ndlxfc2}), (\ref{ndlxfc2a}), and (\ref{ndlxfc2b})
are the form of Newton's motion equation. In this case, although
particle 1 has a force $-\bf F_1$ applied by the testing particle,
particle 1 is an inertial reference system.

[5] From equation (\ref{ndlxfcb}), (\ref{ndlxfc1}), or
(\ref{ndlxfc1a}), we have
\begin{equation}
{\bf F}=\frac{d(m{\bf V})}{dt} =m\frac{d{\bf V}}{dt}+{\bf
V}\frac{dm}{dt}. \label{ndfc}
\end{equation}
If the re-definition of the force $\tilde{\bf F}$ is
\begin{equation}
\tilde{\bf F}:= m\frac{d{\bf V}}{dt}. \label{li}
\end{equation}
then
\begin{equation}
\tilde{\bf F}={\bf F}-{\bf V}\frac{dm}{dt}.
\label{guanxinpaichili}
\end{equation}
When an object is moving to the center of substance, that is
$\frac{dm}{dt}>0$, then $-{\bf V}\frac{dm}{dt}$ and ${\bf V}$ are
in opposite directions. On the contrary, when an object is away
from the center of substance, that is $\frac{dm}{dt}<0$, then
$-{\bf V}\frac{dm}{dt}$ and ${\bf V}$ is in the same direction.
Therefore, the term $-{\bf V}\frac{dm}{dt}$ has a repulsion effect
due to the inertial interaction; that is, the term $-{\bf
V}\frac{dm}{dt}$ becomes a repulsive force.

Perhaps the accelerating expansion of the universe is related to
this. The spaceship flying to or from the Milky Way will have an
additional acceleration. The former will be discussed in $\S 6$
and the latter will be discussed in the following. For simplicity,
let the Earth, the Sun, and the center of the Milky Way be still
relative to each other, on the same straight line, and have the
spaceship move along the line. From equation
(\ref{guanxinpaichili}), we know that the additional acceleration
of the spaceship due to the inertial interaction is

\begin{equation}
\begin{array}{l}
{\bf a}_I=-\frac{1}{m}{\bf V}\frac{dm}{dt}\\
=-\frac{1}{m}{\bf V}\frac{d}{dt}[\alpha_Um_G+K\frac{M_Mm_G}{r_M}+
K\frac{M_{\odot}m_G}{r_{\odot}}+K\frac{M_Em_G}{r_E}]\\
=\frac{1}{m}{\bf V}(K\frac{M_Mm_G}{r_M}\frac{\dot{r}_M}{r_M}+
K\frac{M_{\odot}m_G}{r_{\odot}}\frac{\dot{r}_{\odot}}{r_{\odot}}
+K\frac{M_Em_G}{r_E}\frac{\dot{r}_E}{r_E})\\.
\end{array}
\label{gxjsd}
\end{equation}
here, $m$: the inertial mass of the spaceship; $m_G$: the
gravitational mass of the spaceship; $r_M$, $r_{\odot}$ and $r_E$:
the distances of the spaceship from the centers of the Milky Way,
the Sun, and the Earth. In the following three cases will be
discussed:

(1) Near the Earth, due to
$|\dot{r}_M|=|\dot{r}_{\odot}|=|\dot{r}_E|$,
$\alpha_M\approx10\alpha_{\odot}\approx100\alpha_E$ and
$\alpha_M\frac{1}{r_M}\approx10^{-9}\alpha_{\odot}\frac{1}{r_{\odot}}
\approx10^{-12}\alpha_E\frac{1}{r_E}$, we can have

\begin{equation}
{\bf a}_I\approx\frac{1}{100}\alpha_M\frac{\dot{r}_E^2}{r_E}{\bf
e}_r \label{gxjsdE}
\end{equation}
where ${\bf e}_r$ is the unit vector positioned from the center of
the Milky Way to the spaceship. For example, assuming
$V\equiv\dot{r}_E\sim \pm 10^3ms^{-1}$ and $r_E\sim 10^7m$, then
we have ${\bf a}_I\sim 8.2\times 10^{-5}ms^{-2}{\bf e}_r$.

(2)The spaceship is in the interior of solar system, but away from
the Earth. Therefore, we have $\frac{M_{\odot}}{r_{\odot}^2} \gg
\frac{M_E}{r_E^2}$ ($r_E>10^{-2}r_{\odot}$ is enough). Then
\begin{equation}
{\bf
a}_I\approx\frac{1}{10}\alpha_M\frac{\dot{r}_{\odot}^2}{r_{\odot}}{\bf
e}_r \label{gxjsdS}
\end{equation}
For example, assuming $V\equiv\dot{r}_{\odot}\sim \pm 10^4ms^{-1}$
and $r_{\odot}\sim 10^{11}m$, then we have ${\bf a}_I\sim
8.2\times 10^{-6}ms^{-2}{\bf e}_r$.

(3)The spaceship is moving into the Milky Way, but away from solar
system. Therefore, we have $\frac{M_M}{r_M^2} \gg
\frac{M_{\odot}}{r_{\odot}^2} \gg \frac{M_E}{r_E^2}$
($r_\odot>10^{-5}r_M$ is enough). Then
\begin{equation}
{\bf a}_I\approx \alpha_M\frac{\dot{r}_M^2}{r_M}{\bf e}_r
\label{gxjsdM}
\end{equation}
For example, assuming $V\equiv\dot{r}_M\sim \pm 10^5ms^{-1}$ and
$r_M\sim 10^{20}m$, then we have ${\bf a}_I\sim 8.2\times
10^{-11}ms^{-2}{\bf e}_r$.

\subsection{THE MOTION EQUATION AND THE ENERGY FORMULA OF AN OBJECT IN MINKOWSKI'S SPACETIME}

In Minkowski's spacetime $(M,\eta_{ab})$, let ${\tilde m}_G$ be
the gravitational mass of the testing particle and $m$ be the
inertial mass of the testing particle by the whole of substances
in the spacetime other than itself through the inertial
interaction. From the formula of inertial mass, $m$ can be
expressed as follows
\begin{equation}
m=f(P){\tilde m}_G. \label{gunaxingzhiliang}
\end{equation}
where, $P\in M$ is the spacetime point in which the testing
particle is located, and $f(P)$ is the function of the space-time
point that is determined by the inertial interaction of all the
substances, excluding the testing particle itself, in the
spacetime. $f(P)$ has nothing to do with ${\tilde m}_G$.

$m=\gamma m_0$ is seen by the special relativity. $m_0$ is the
still inertial mass,
$\gamma\equiv(1-\frac{u^2}{c^2})^{-\frac{1}{2}}$, while $u$ is the
3-velocity of the testing particle relative to the observer. On
the Earth, we have $m={\tilde m}_G$ and $m_0=m_G$. $m_G$ is the
still gravitational mass of the testing particle, then we have
${\tilde m}_G=\gamma m_G$. It can be assumed that ${\tilde
m}_G=\gamma m_G$ is generally applicable. Therefore,
\begin{equation}
m=f(P){\tilde m}_G=f(P)\gamma m_G. \label{gunaxingzhiliang1}
\end{equation}
Perhaps, $m_G$ depends on the intrinsic properties and the
fundamental interactions including or except for inertial
interactions of the particles which have no structure and compose
the testing particle (i.e., elementary particles). While $\gamma$
is decided by the basic interactions outside of the testing
particle in addition to the inertial interaction, $f(P)$ is
determined by the inertial interaction outside the testing
particle.

\subsubsection{THE TESTING PARTICLE IS ELEMENTARY PARTICLE}

4-force $F^a$ of the testing particle can be defined as
\begin{equation}
F^a:=U^b\partial_b (f(P)m_GU^a). \label{msdlxfc}
\end{equation}
where, $\partial_a$ is the derivative operator associated with
Minkowski's metric $\eta_{ab}$. $U^a$ is the 4-velocity of the
testing particle. $m_G$ is the still gravitational mass of the
testing particle which has no structure. $m_G$ is the intrinsic
property of elementary particle. Of course, here we do not discuss
quantum mechanics.

If there exists an inertial frame, meaning that all objects except
the testing particle are relatively static, then we can assume
that the motion equation of the testing particle for this
coordinate system is
\begin{equation}
{\bf F}=\frac{d}{dt}(f(P)\gamma m_G{\bf u}). \label{3lifc}
\end{equation}
where, $\bf u$ and $\bf F$ are respectively the 3-velocity and
3-force of the testing particle. $t$ is the coordinate time of the
inertial frame.

The following equations can be proved
\begin{equation}
F^i=\gamma f^i. \label{4li3}
\end{equation}
\begin{equation}
F^0=\gamma {\bf f}\cdot{\bf u}+c^2\gamma m_G\frac{d}{dt}f(P).
\label{4li0}
\end{equation}

Please note the following: [1] When the relationships (\ref{4li3})
and (\ref{4li0}) between 4-force $F^a$ and 3-force $\bf f$ are
known ,equation (\ref{msdlxfc}) also can be called the motion
equation of the testing particle. [2]If there is no any inertial
frame, the relationship between 4-force and 3-force needs to be
further studied; [3]For a free particle ${\bf f}=0$, we have
$F^i=0$, yet $F^0=c^2\gamma m_G\frac{d}{dt}f(P)\not=0$. From
equation (\ref{msdlxfc}),we also know that 4-force $F^a\not=0$ of
the testing particle due to the inertial interaction.

From equation (\ref{3lifc}), we have that from $t_1$ to $t_2$,
\begin{equation}
\int_{t_1}^{t_2}{\bf F}\cdot{\bf
u}dt+\int_{P_1}^{P_2}c\sqrt{c^2-u^2}m_G df(P)=m_2 c^2-m_1c^2.
\label{nengliangfc}
\end{equation}
where, $m_1=f(P_1)\gamma_1 m_G$ and $m_2=f(P_2)\gamma_2 m_G$ are
respectively the inertial masses of the testing particle at $t_1$
and $t_2$. Like the special relativity theory, the energy of the
elementary particles is defined as
\begin{equation}
E=mc^2=f(P)\gamma m_Gc^2. \label{nengliang0}
\end{equation}
If the Higgs mechanism really does exist, the mass which the
particle obtains by the Higgs mechanism is likely to be the
intrinsic gravitational mass $m_G$ of the elementary particle.

\subsubsection{TESTING PARTICLE HAVING INTERNAL STRUCTURE}

For the testing particle which has internal structure, there are
three possibilities:

1)Assume the motion equation of the testing particle be as
follows(That is the definition formula of 4-force)
\begin{equation}
F^a:=U^b\partial_b [(f(P)m_G+m_{in})U^a]. \label{msdlxfc1}
\end{equation}
where, $m_G$ is the rest gravitational mass of testing particle,
which is determined by the intrinsic gravitational mass of
elementary particles composing the testing particle and the
fundamental interactions except for inertial interactions of the
elementary particles. $f(P)$ is decided by the inertial
interaction outside of the testing particle. It is noticed that
$m_{in}$ is the internal inertial mass of the testing particle
rather than the bound energy of the testing particle, and the
bound energy of the testing particle is reflected in $m_G$.

It is can be imagined that the testing particle is formed together
by elementary particles from the dispersed state without inertial
interaction. Then, it is reasonable to assume that the energy of
the testing particle is
\begin{equation}
E=\gamma [f(P)m_G+m_{in}]c^2=E_I+E_{in}. \label{nengliang1}
\end{equation}
here, $E_I\equiv \gamma f(P)m_Gc^2$ and $E_{in}\equiv \gamma
m_{in}c^2$, $\gamma$ is decided by the basic interactions outside
the testing particle except for the inertial interaction.

2)Assume the motion equation and the energy of the testing
particle be respectively
\begin{equation}
F^a:=U^b\partial_b [f(P)m_GU^a]. \label{msdlxfc2}
\end{equation}
\begin{equation}
E=\gamma [f(P)m_G+m_{in}]c^2=E_I+E_{in}. \label{nengliang2}
\end{equation}
where, $E_I\equiv \gamma f(P)m_Gc^2$ and $E_{in}\equiv \gamma
m_{in}c^2$.

3)Assume the motion equation and the energy of the testing
particle be respectively
\begin{equation}
F^a:=U^b\partial_b [f(P)m_GU^a]. \label{msdlxfc3}
\end{equation}
\begin{equation}
E=E_I\equiv \gamma f(P)m_Gc^2. \label{nengliang3}
\end{equation}
where, $m_G$ is the gravitational mass of testing particle, which
is determined by the intrinsic gravitational mass of elementary
particles composing the testing particle and the fundamental
interactions including inertial interactions of the elementary
particles. The internal inertial interactions of the testing
particle are reflected in the effects for $m_G$.

Please note the following:[1]The relationship between 4 force and
3 force needs to be further studied.[2] If there are only two
particles in Minkowski's spacetime, then every particle is an
inertial frame for the other particle. But the world lines are not
geodesic due to the interaction force between the two. If the
motion equation of the testing particle is the case 2) or 3),
because the motion of the two particles is completely symmetrical,
the proper time of the two particles are the same after the
separation of the two particles and then to meet again. If the
motion equation of the testing particle is case 1), because the
motion of the two particles is not symmetrical, the proper time of
the two particles are not the same after the separation of the two
particles and then to meet again. Generally there does not exist
inertial frame. The reference frame is an inertial frame only when
there are no relative motions of all viewers in reference frame.
In special relativity, describing the inertial system by Lorenzian
coordinate system is just an approximate case when the force of
the inertial frame by the testing particle can be ignored. In case
1), this approximation, after all, is a good approximation.
However, in case 2) and 3), ignoring the force of the inertial
frame by the testing particle is not very plausible. This
treatment can only be a practical way.

\subsection{THE MOTION EQUATION OF AN OBJECT IN A GENERAL SPACETIME}

Let $(M,g_{ab})$ be a general space-time

If the testing particle is an elementary particle, it is assumed
that the motion equation (That is the definition formula of
4-force) and the energy of the testing particle be respectively
\begin{equation}
F^a:=U^b\nabla_b (f(P)m_GU^a). \label{ryskdlxfc}
\end{equation}
\begin{equation}
E=mc^2=f(P)\gamma m_Gc^2. \label{rysknengliang}
\end{equation}
where, $\nabla_a$ is the derivative operator associated with the
space-time's metric $g_{ab}$. $F^a$ and $U^a$ are respectively the
4-force and 4-velocity of the testing particle.

For the testing particle which have internal structure, there are
also three possibilities corresponding to Minkowski's spacetime:

(1)Assume the motion equation (That is the definition formula of
4-force) and the energy of the testing particle be respectively
\begin{equation}
F^a:=U^b\nabla_b [(f(P)m_G+m_{in})U^a]. \label{ryskdlxfc1}
\end{equation}
\begin{equation}
E=\gamma [f(P)m_G+m_{in}]c^2=E_I+E_{in}. \label{rysknengliang1}
\end{equation}
where, $E_I\equiv \gamma f(P)m_Gc^2$ and $E_{in}\equiv \gamma
m_{in}c^2$

(2)Assume the motion equation and the energy of the testing
particle be respectively

\begin{equation}
F^a:=U^b\nabla_b [f(P)m_GU^a]. \label{ryskdlxfc2}
\end{equation}
\begin{equation}
E=\gamma [f(P)m_G+m_{in}]c^2=E_I+E_{in}. \label{rysknengliang2}
\end{equation}
where, $E_I\equiv \gamma f(P)m_Gc^2$, and $E_{in}\equiv \gamma
m_{in}c^2$.

(3)Assume the motion equation and the energy of the testing
particle be respectively
\begin{equation}
F^a:=U^b\nabla_b [f(P)m_GU^a]. \label{ryskdlxfc3}
\end{equation}
\begin{equation}
E_I=\gamma f(P)m_Gc^2. \label{rysknengliang3}
\end{equation}

Please note:

1)Because the inertial interaction is different from the known
four fundamental interactions, the inertial interaction directly
impacts the energy of objects (this is reflected in $f(P)$ and
$m_{in}$ of the energy expression). The inertial interaction also
produces the inertial force in equations (\ref{ndlxfc}) and
(\ref{ndlxfca}) and the repulsive force in equation
(\ref{guanxinpaichili}). Therefore perhaps the inertial
interaction is a new interaction. Or at least it is another aspect
of the gravitational interaction which people have not recognized
yet.

2) From equation
(\ref{ryskdlxfc}),(\ref{ryskdlxfc1}),(\ref{ryskdlxfc2}), and
(\ref{ryskdlxfc3}),we know that 4-force $F^a\not=0$ of the testing
particle due to the inertial interaction. The relationship between
4 force and 3 force needs to be further studied.Which one of the
above three possibilities, or some other case, is the particle
motion equation and energy should be determined by experiment and
observation.

3)In the case (2) and (3), $\frac{\gamma f(P)m_G}{\gamma
m_G}=f(P)$, the ratio of the inertial mass and the gravitational
mass of an object is dependent on the point in space-time, instead
of independent objects. In the case (1), $\frac{\gamma
(f(P)m_G+m_{in})}{\gamma m_G}=f(P)+\frac{m_{in}}{m_G}$, the ratio
of the inertial mass and the gravitational mass of an object is
dependent on object. However, if $m_{in}\ll m_G$, $\frac{\gamma
(f(P)m_G+m_{in})}{\gamma m_G}\approx f(P)$ is independent on
object.

4)Assuming that $L(\tau)$ is the world line of any particle,
$\tau$ is the proper time of the particle, $U^a$ is the
4-velocity, and $P\in L$, then the 3-velocity of the particle
relative to any instantaneous observer $(P,Z^a)$ ($Z^a$ is the
4-velocity the observer) can be defined as
\begin{equation}
u^a:={h^a}_bU^b/\gamma . \label{3su}
\end{equation}
where, ${h^a}_b\equiv g^{ac}h_{cb}$, $h_{ab}\equiv g_{ab}+Z_aZ_b$,
and $\gamma \equiv -U^aZ_a$.

The 3-speed of a particle relative to any instantaneous observer
$(P,Z^a)$ can be defined as
\begin{equation}
u:=\sqrt{u^au_b}. \label{3sulv}
\end{equation}
where, $u_a:=h_{ab}u^b$. It can be proved that
(1)$\gamma=(1-\frac{u^2}{c^2})^{-\frac{1}{2}}$;
(2)$U^a=\gamma(Z^a+u^a)$.

5)The 4-momentum of a particle is defined as
\begin{equation}
P^a:=f(P)m_GU^a. \label{4dongliang}
\end{equation}
The 4-momentum of a particle can be decomposed as 3+1 by an
instantaneous observer $(P,Z^a)$.
\begin{equation}
P^a=E_IZ^a+p^a. \label{4dongliang3+1}
\end{equation}
where, $p^a\equiv \gamma f(P)m_Gu^a$ is the 3-momentum of a
particle relative to an instantaneous observer $(P,Z^a)$,
obviously, $E_I=-P^aZ_a$.

Or the 4-momentum of a particle is defined as
\begin{equation}
P^a:=(f(P)m_G+m_{in})U^a. \label{4dongliang1}
\end{equation}
The 4-momentum of a particle can be decomposed as 3+1 by an
instantaneous observer $(P,Z^a)$

\begin{equation}
P^a=EZ^a+p^a. \label{4dongliang13+1}
\end{equation}
where, $p^a\equiv \gamma (f(P)m_G+m_{in})u^a$ is the 3-momentum of
a particle relative to an instantaneous observer $(P,Z^a)$,
obviously, $E=-P^aZ_a$.

\section{THEORY OF GRAVITY}

\subsection{EINSTEIN'S THEORY OF GRAVITY PERHAPS NEEDS TO BE MODIFIED}

Einstein's theory of gravity perhaps needs to be modified. At
least, there are three reasons. First, if the inertial interaction
does exist, $T_{ab}$, the energy-momentum tensor field of
substances does not satisfy $\nabla^aT_{ab}=0$; but Einstein
tensor $G_{ab}$ satisfies $\nabla^aG_{ab}=0$. For example, the
inertial mass of a given system, though without exchange of matter
with the outside, will still be changed, because the outside
matter changes distribution. It means that the continuity equation
of inertial mass is not satisfied. Second, taking any cross
section in the perfect fluid, the matters which lie on each side
of the cross-section generate attractive effect due to the
inertial mass density $\rho$ and exclusion effect due to the
pressure $p$. But in Einstein's equation, both the pressure $p>0$
and the inertial mass density $\rho$ generate attractive effect;
only the pressure $p<0$ generates exclusion effect. This can be
found from Einstein's equation of the perfect fluid
\begin{equation}
R_{ab}-\frac{1}{2}Rg_{ab}=8\pi[(\rho+p)U_aU_b+pg_{ab}].
\label{eystylcfc}
\end{equation}
The pressure $p$ not only gives to the contribution of the second
term, but also it appears in the first item, with the same sign as
$\rho$, in the "source" of the right hand side in equation
(\ref{eystylcfc}). Both $p$ and $\rho$ produce attractive effect.
As a concrete example, the above can be seen from one of the
Einstein's cosmic evolution equations\cite{liangchanbin},
$3\ddot{a}=-4\pi a(\rho+3p)$. Third, the Einstein's equation does
not imply the inertial exclusion effect reflected in equation
(\ref{guanxinpaichili}).

The inertial mass and the gravitational mass of an object are
generally not equal due to the inertial interaction. It means that
the equivalence principle about the equivalence of the inertial
mass and the gravitational mass of an object perhaps does not
hold. From equations (\ref{ryskdlxfc2}),(\ref{ryskdlxfc3}), and
(\ref{ryskdlxfc1}) we know that the ratio of inertial mass and
gravitational mass, $\frac{\gamma f(P)m_G}{\gamma m_G}=f(P)$ or
$\frac{\gamma (f(P)m_G+m_{in})}{\gamma m_G}\approx f(P)$, is
spacetime point dependent, but non-object dependent or
approximately non-object dependent. It also means that the ratio
of inertial mass and gravitational mass of all objects, measured
at the same point in spacetime, is the same. However, if measured
at different points in spacetime, the ratio is different. "The
ratio of inertial mass and gravitational mass of all objects,
measured at the same point in spacetime, is the same" can be
regarded as the generalized equivalence principle about the
equivalence of the inertial mass and the gravitational mass.

Due to the generalized equivalence principle about the equivalence
of the inertial mass and the gravitational mass, the world lines
of free particles are non-object dependent. We also can assume
that the world line of a free particle is geodesic. Therefore, the
theory of gravity should be still the theory about geometry. But
if the motion equation of a particle with internal structure is
equation (\ref{ryskdlxfc1}), the description to a particle with
internal structure is approximate in the geometry theory of
gravity .

\subsection{TIDAL PHENOMENON}

\subsubsection{THE TIDAL PHENOMENON OF NEWTON'S THEORY OF GRAVITY IN NEWTON'S SPACETIME}

In Newton's space-time, the relationship between the inertial mass
$m$ and the gravitational mass $m_G$ is as follows.

\begin{equation}
m=f({\bf r})m_G. \label{gunaxingzhiliang2}
\end{equation}
where, $f({\bf r})$ is the function of a point in space which is
determined by the inertial interaction of all the substances in
the spacetime other than the testing particle itself. $f({\bf r})$
has nothing to do with $m_G$. When the motion equation is equation
(\ref{ndlxfc1a}) and if $m_{in}$ can be ignored, then equation
(\ref{gunaxingzhiliang2}) applies approximately. In a small
spatial extent, the variation of $f({\bf r})$ with respect to $\bf
r$ can be ignored. It is an approximation that $f({\bf r})$ does
not change with $\bf r$.

Let ${\bf r}(t)\equiv x^i(t){\bf e}_i$ and ${\bf r}(t)+{\bf
w}(t)\equiv [x^i(t)+w^i(t)]{\bf e}_i$ be respectively the spatial
position vector of adjacent particle 1 and 2 free-falling in the
gravitational field. ${\bf e}_i$ is the basis of Cartesian
coordinate system. Then ${\bf w}(t)\equiv w^i(t){\bf e}_i$ is the
position vector of particle 2 relative to particle 1.
$\frac{d^2{\bf w}}{dt^2}$ is the tidal acceleration of particle 2
relative to particle 1. Let $\phi$ be Newtonian gravitational
potential. Then from the motion equations (\ref{ndlxfc1}) or
(\ref{ndlxfc1a}) and Newton's law of universal gravitation, we can
obtain
\begin{equation}
f({\bf r})\frac{d^2w^i}{dt^2}\approx
-\frac{\partial^2\phi}{\partial x^j\partial x^i}|_{\bf r}w^j.
\label{ndcxfc}
\end{equation}

\subsubsection{THE TIDAL PHENOMENON IN GENERAL SPACETIME}

Let $(M,g_{ab})$ be a general spacetime. Let $\gamma_s(\tau)$
denote a smooth one-parameter family of the world lines of free
particles in gravitational field and $\Sigma$ denote the
two-dimensional submanifold spanned by the curves
$\gamma_s(\tau)$. $(\tau,s)$ can be the coordinates of $\Sigma$,
and $[Z,w]^a=0$, where,
$Z^a\equiv(\frac{\partial}{\partial\tau})^a$ and
$w^a\equiv(\frac{\partial}{\partial s})^a$. The relative velocity
and the relative acceleration of infinitesimally nearby free
particles can be respectively defined as
\begin{equation}
u^a=Z^b\nabla_b w^a. \label{xdsd}
\end{equation}
and
\begin{equation}
a^a=Z^b\nabla_b u^a. \label{xdjsd}
\end{equation}
The following geodesic deviation equation can be proved:
\begin{equation}
a^c=-{R_{abd}}^cZ^aw^bZ^d.
\label{cdplfc}
\end{equation}

\subsection{THE EQUATION OF GRAVITATIONAL FIELD}

Equation (\ref{ndcxfc}) can be rewritten as
\begin{equation}
a^c=(\frac{\partial}{\partial x^i})^c\frac{d^2w^i}{dt^2}\approx
-\frac{1}{f({\bf r})}w^b\partial_b\partial^c \phi. \label{ndcxfc1}
\end{equation}
The comparison of equation (\ref{cdplfc}) and (\ref{ndcxfc1})
implies the following correspondence
\begin{equation}
{R_{abd}}^cZ^aZ^d\longleftrightarrow \frac{1}{f({\bf
r})}\partial_b\partial^c \phi. \label{cdplndcxfc}
\end{equation}
Contracting the index $c$ and $b$, we have
\begin{equation}
R_{ad}Z^aZ^d\longleftrightarrow \frac{1}{f({\bf
r})}\partial_b\partial^b \phi= \frac{1}{f({\bf r})}\nabla^2\phi
=4\pi\frac{1}{f({\bf r})}\rho_G =\frac{1}{f^2({\bf r})}4\pi\rho.
\label{cdplndcxfc1}
\end{equation}
where, $\rho_G$ and $\rho=f({\bf r})\rho_G$ are respectively the
gravitational mass density and the inertial mass density of
matter.

From the discussion in the beginning of $\S 4.1$, we have known
that if there does exist inertial interaction then it is not the
energy-momentum tensor $T_{ab}$ to decide the geometry properties
of spacetime.

Definition: the tensor field deciding the geometry properties of
spacetime is called a "matter field tensor" , denoted as $M_{ab}$.

From  equation (\ref{cdplndcxfc1}) it can be assumed that $M_{ab}$
satisfies
\begin{equation}
\rho=M_{ab}Z^aZ^b \label{Mab}
\end{equation}
$\rho$ is the inertial mass density of matter measured by an
observer $Z^a$. Then equation (\ref{cdplndcxfc1}) becomes

\begin{equation}
R_{ad}Z^aZ^d\longleftrightarrow \frac{1}{f^2({\bf r})}4\pi
M_{ab}Z^aZ^b. \label{cdplndcxfc2}
\end{equation}

Please note the following

(1)Assume the matter field tensor of a perfect fluid be
$M_{ab}=(\rho+f_1(p,\rho))U_aU_b +\tilde f_1(p,\rho)g_{ab}$, and
we might as well take $f_1(p,\rho))=\tilde f_1(p,\rho)$, then
\begin{equation}
M_{ab}=(\rho+f_1(p,\rho))U_aU_b+f_1(p,\rho)g_{ab}.
\label{lxltwzczl}
\end{equation}
where, $U^a=g^{ab}U_b$ is the 4-velocity field of perfect fluid.
$\rho=M_{ab}U^aU^b$ is the inertial mass density of perfect fluid
measured by a co-moving observer. $p$ is the pressure of perfect
fluid.
\begin{equation}
f_1(p,\rho)\equiv \hat{\beta} p+f_2(\rho). \label{f_1}
\end{equation}
$\hat{\beta}$ is a constant to be determined. Its value is
discussed later. $f_2(\rho)$ is related to the change rate with
proper time of the inertial mass and the speed of movement of
perfect fluid's matter element, which will be discussed in the
cosmology portion.

(2)The matter field tensor $M_{ab}$ is different from the
energy-momentum tensor $T_{ab}$, but both $p$ and $f_2(\rho)$ can
be ignored for the non-relativistic dust matter within the small
scope of space. Then $M_{ab}\approx T_{ab}$.

(3) $\rho=\lim\limits_{\bigtriangleup V\to 0} \frac{\bigtriangleup
m}{\bigtriangleup V}$ in $M_{ab}$, $\bigtriangleup V$ is the
volume element. $\bigtriangleup m$ is the inertial mass in
$\bigtriangleup V$. Then $\rho$ in $M_{ab}$ contains $m_{in}$ in
equation (\ref{rysknengliang1}) or (\ref{rysknengliang2}).

(4) Due to the Lorentz Covarience of vacuum, i.e., the status of
vacuum has nothing to do with the observer. It is reasonable to
assume that vacuum has no inertial mass and gravitational mass.
That is, assuming that the energy of vacuum is not included in the
energy of equations (\ref{rysknengliang1}),
(\ref{rysknengliang2}), and (\ref{rysknengliang3}), therefore,
neither is it included in the matter field tensor $M_{ab}$. The
energy of vacuum does not affect the property of spacetime.
Furthermore, the Casimir force $F_{cas}\propto \frac{1}{d^4}$, $d$
is the distance between the two flat conductors. When
$d\rightarrow \infty$, then $F_{cas}\rightarrow 0$. It is
reasonable to assume that the pressure of infinite vacuum is
$p=0$. Therefore, it is reasonable to assume that vacuum does not
affect the property of spacetime.

(5)The need of negative pressure in the Einstein's theory may be
due to the attractive effect of the pressure in the Einstein's
theory. Because $p<0$ appears only in the theoretical study of
vacuum, cosmological constant, and the dark energy, the physical
phenomenon of $p<0$ is not directly observed and the physical
image and physical essence of $p<0$ is not clear. We assume $p\geq
0$ in this paper.

From equation (\ref{cdplndcxfc2}), assume the equation of
gravitational field be
\begin{equation}
R_{ab}+\alpha Rg_{ab}+\Lambda g_{ab}=\tilde{\kappa}M_{ab}.
\label{ylcfc}
\end{equation}
where, $\tilde{\kappa}\equiv \frac{\kappa}{f^2(P)}$, $\alpha$, and
$\kappa$ are constants to be determined. $\Lambda$ is the
cosmological constant.

From equation (\ref{ylcfc}), we have
\begin{equation}
R=\frac{\tilde{\kappa}M-4\Lambda}{4\alpha+1}. \label{R}
\end{equation}
where, $M\equiv {M_a}^a$.
For a perfect fluid and any observer whose 4-velocity is $Z^a$, we
have
\begin{equation}
R_{ab}Z^aZ^b=\frac{\gamma^2(4\alpha+1)-\alpha}{4\alpha+1}\tilde{\kappa}\rho
+\tilde{\kappa}(\gamma^2-\frac{\alpha+1}{4\alpha+1})f_1(p,\rho)
+\frac{\Lambda}{4\alpha+1}. \label{R_{ab}Z^aZ^b} \end{equation}
where, $\gamma\equiv -U^aZ_a$. It is easy to prove that the
energy-momentum tensor $T_{ab}$ of the perfect fluid of $\rho\geq
0$ and $p\geq 0$ satisfies weak energy condition, strong energy
condition and dominant energy condition. To avoid space-time
singularity in our theory as much as possible, it is required that
there is an observer $Z^a$ so that in an evolutionary stage of the
observer $R_{ab}Z^aZ^b<0$ satisfies. It means that the conditions
of singularity theorem are not satisfied.

For the dust matter within the small scope of space, when
$\Lambda$ can be ignored and Newtonian approximation applies,
$\gamma\equiv -U^aZ_a\approx 1$ and $f_1(p,\rho)\approx 0$, we
have

\begin{equation}
R_{ab}Z^aZ^b\approx\frac{3\alpha+1}{4\alpha+1}\tilde{\kappa}\rho.
\label{R_{ab}Z^aZ^b1}
\end{equation}
Compared with the equation (\ref{cdplndcxfc1}), it can be obtained
\begin{equation} \kappa\frac{3\alpha+1}{4\alpha+1}=4\pi.
\label{kappa}
\end{equation}
From equation (\ref{kappa}), we know $\alpha\neq -\frac{1}{3}$ and
$\alpha\neq -\frac{1}{4}$.

\subsection{THE EQUATION OF GRAVITATIONAL FIELD IN VACUUM}

For vacuum $M_{ab}=0$, ignoring $\Lambda$, the equation of
gravitation (\ref{ylcfc}) becomes
\begin{equation}
R_{ab}=0. \label{zkylcfc}
\end{equation}
It is the same as the Einstein's vacuum equation in form.

\subsection{LINEARIZED THEORY OF GRAVITY AND THE NEWTONIAN LIMIT}
\subsubsection{LINEARIZED THEORY OF GRAVITY}

In the case of a weak gravitational field, the metric $g_{ab}$ of
spacetime is very close to Minkowski metric $\eta_{ab}$.
$\gamma_{ab}$ can be defined as
\begin{equation}
g_{ab}=\eta_{ab}+\gamma_{ab}. \label{gamma_{ab}}
\end{equation}
$\gamma_{ab}$ is very small. Substituting equation
(\ref{gamma_{ab}}) into the gravitational field equation
(\ref{ylcfc}), ignoring $\Lambda$ and taking first order
approximation, the linear gravitational field equation can be
obtained
\begin{equation}
\partial^c\partial_{(a}\gamma_{b)c}-\frac{1}{2}\partial^c\partial_c\gamma_{ab}
-\frac{1}{2}\partial_a\partial_b\gamma+\alpha \eta_{ab}
(\partial^c\partial^d\gamma_{cd}-\partial^c\partial_c\gamma)
=\tilde{\kappa}M_{ab}. \label{xianxingylcfc}
\end{equation}
The equation (\ref{xianxingylcfc}) is the linearized equation of
$\gamma_{ab}$ and the linear approximation of the gravitational
field equations.

There is a gauge freedom in any theory in which gravity is
described by metric\cite{liangchanbin}\cite{wald}. Let $\phi :M\to
M$ be a diffeomorphism, then both $(M,g_{ab})$ and
$(M,\phi^{\ast}g_{ab})$ represent the same physical spacetime.
Thus, we can always find the vector field $\xi^a=\eta^{ab}\xi_b$.
For the gauge transformation
\begin{equation}
\tilde{\gamma}_{cd}=\gamma_{cd}+\partial_a\xi_b+\partial_b\xi_a.
\label{gfbh}
\end{equation}
Let $00$ component of equation (\ref{xianxingylcfc})) in global
inertial coordinate system\cite{wald} be divided into two
equations
\begin{equation}
\partial^c\partial_c\gamma_{00}=-\frac{\sigma}{f^2(P)}M_{00}.
\label{gamma00}
\end{equation}
\begin{equation}
\partial^c\partial_{(0}\gamma_{0)c}
-\frac{1}{2}\partial_0\partial_0\gamma+\alpha \eta_{00}
(\partial^c\partial^d\gamma_{cd}-\partial^c\partial_c\gamma)
=(\tilde{\kappa}-\frac{\sigma}{2f^2(P)})M_{00}. \label{gamma00a}
\end{equation}
where $\sigma$ is a constant to be determined. In fact, the 15
unknown functions $\gamma_{\mu\nu}$, $U^{\mu}$ and $\rho$ (in
Newton approximation, $p\approx 0$) can be obtained so that the 12
equations (\ref{xianxingylcfc}), (\ref{gamma00}), (\ref{gamma00a})
and $g_{ab}U^aU^b=-1$ are satisfied.

\subsubsection{THE NEWTONIAN LIMIT}

From the following discussion we can see that Newton's theory of
gravity can be regarded as a limiting case of the gravitational
theory in the weak field under low-speed conditions.

(1)Weak field means equation (\ref{gamma_{ab}}).

(2)\texttt{}Low speed means that the field source and the objects
move in low-speed. The low-speed movement of the field source
leads to the slow change of spacetime geometry, $\frac{\partial
\gamma_{\mu\nu}}{\partial t}\approx 0$. The low-speed movement of
objects also leads to that the 4-velocity of objects
$U^a=(\frac{\partial }{\partial \tau})^a$ approximately equals to
the 4-velocity of inertial observer $Z^a=(\frac{\partial
}{\partial t})^a$ of the inertial coordinate system $\{t,x^i\}$.
That is, the proper time $\tau$ of the object is approximately
equal to the coordinate time $t$, $\tau\approx t$. From equation
(\ref{Mab}), we have
\begin{equation}
\rho=M_{ab}U^aU^b\approx M_{ab}Z^aZ^b=M_{00} \label{M00}
\end{equation}
Substituting equation (\ref{M00}) into equation (\ref{gamma00}),
we have
\begin{equation}
\nabla^2\gamma_{00}=-\frac{\sigma}{f^2(P)}\rho=-\frac{\sigma}{f(P)}\rho_G.
\label{gamma00b}
\end{equation}
As an approximation, ignoring the change of $f(P)$ with the point
of space-time, equation (\ref{gamma00b}) becomes
\begin{equation}
\nabla^2(-\frac{4\pi f(P)}{\sigma}\gamma_{00})\approx 4\pi\rho_G.
\label{gamma00c}
\end{equation}
Comparing with Newtonian's gravity equation
\begin{equation}
\nabla^2\phi=4\pi\rho_G. \label{ndylfc}
\end{equation}
we have
\begin{equation}
\phi\approx -\frac{4\pi}{\sigma}f(P)\gamma_{00}. \label{phi}
\end{equation}
The world line of a free particle in the gravitational field is
geodesic.In inertial coordinate system $\{t,x^i\}$, the geodesic
can be expressed as
\begin{equation}
\frac{d^2x^{\mu}}{d\tau^2}+{\Gamma^{\mu}}_{\nu\sigma}
\frac{dx^{\nu}}{d\tau}\frac{dx^{\sigma}}{d\tau}=0. \label{cdx}
\end{equation}
where $\tau$ is the affine parameter of the geodesic.  In the case
of low-speed, $\frac{dx^i}{d\tau}=\frac{dx^i}{dt}\approx 0$,
equation (\ref{cdx}) becomes
\begin{equation}
\frac{d^2x^{\mu}}{dt^2}=-{\Gamma^{\mu}}_{00}.
\label{cdx1}
\end{equation}
\begin{equation}
{\Gamma^i}_{00}=\frac{1}{2}\eta^{ij}(\gamma_{j0,0}+\gamma_{0j,0}-\gamma_{00,j})
\approx -\frac{1}{2}\frac{\partial \gamma_{00}}{\partial x^i}
\approx \frac{\sigma}{8\pi f(P)}\frac{\partial \phi}{\partial
x^i}. \label{cdx3}
\end{equation}
Substituting equation (\ref{cdx3}) into equation (\ref{cdx1}), we
have
\begin{equation}
\frac{d}{dt}[M_Gf(P)\frac{dx^i}{dt}] \approx
-\frac{\sigma}{8\pi}\frac{\partial (M_G\phi)}{\partial x^i}.
\label{cdx5}
\end{equation}
Comparing with Newtonian's gravity equation
\begin{equation}
\frac{d}{dt}[M_Gf(P)\frac{dx^i}{dt}] =-\frac{\partial
(M_G\phi)}{\partial x^i}. \label{cdx6}
\end{equation}
we have
\begin{equation}
\sigma=8\pi. \label{sigma}
\end{equation}
It is noticed that if let $\sigma=8\pi$, Newton's theory of
gravitation can be regarded as the weak gravitational field and
the limit of low speed of our gravitation theory. Meanwhile, we
have
\begin{equation}
\phi\approx -\frac{1}{2}f(P)\gamma_{00}. \label{phi1}
\end{equation}

\section{STATIC SPHERICALLY SYMMETRIC METRIC}

In static spherically symmetric spacetime $(M,g_{ab})$, the line
element of metric has the following form\cite{liangchanbin}
\begin{equation}
ds^2=-e^{2A(r)}dt^2+e^{2B(r)}dr^2+r^2(d\theta^2+sin^2\theta
d\varphi^2). \label{jtqdcxy}
\end{equation}
where, $A(r)$ and $B(r)$ are the functions with respect to $r$ to
be determined.

\subsection{SCHWARZSCHILD VACUUM SOLUTION}

\subsubsection{SCHWARZSCHILD VACUUM SOLUTION}

Because the vacuum gravitational field equations (\ref{zkylcfc})
is the same as the Einstein's vacuum gravitational field
equations, the static spherically symmetric vacuum metric still
has the following form\cite{liangchanbin}
\begin{equation}
ds^2=-(1+\frac{C}{r})dt^2+(1+\frac{C}{r})^{-1}dr^2+r^2(d\theta^2+sin^2\theta
d\varphi^2). \label{jtqdczkdg}
\end{equation}
Where $C$ is a constant to be determined. By equation (\ref{phi1})
and the Newtonian gravitational potential $\phi=-\frac{M_G}{r}$
and ignoring the changes of $f(P)$ with point in spacetime, we can
get $g_{00}=-1-\frac{C}{r}=
\eta_{00}+\gamma_{00}=-1-\frac{2}{f(P)}\phi=-1-(-\frac{2M_G}{f(P)})\frac{1}{r}$
then $C=-\frac{2M_G}{f(P)}$. Therefore, the static spherically
symmetric vacuum metric is
\begin{equation}
ds^2=-(1-\frac{2M_G}{r}\frac{1}{f(P)})dt^2
+(1-\frac{2M_G}{r}\frac{1}{f(P)})^{-1}dr^2+r^2(d\theta^2+sin^2\theta
d\varphi^2). \label{jtqdczkdg1}
\end{equation}
Namely it is the line element of Schwarzschild spacetime. Please
note that $f(P)$ is regarded as a constant in the range of
spacetime discussed.

In the vacuum region of uniform spherical shell, we have $C=0$ in
equation (\ref{jtqdczkdg}). Otherwise, $g_{00}$ and $g^{11}$ at
$r=0$ will diverge. Then the metric in the vacuum region of
uniform spherical shell is Minkowski metric.

\subsubsection{THE EQUATIONS OF ELECTROMAGNETIC FIELD}

For electromagnetic field, we assume

(1)Electromagnetic field is described by $F_{ab}$ which satisfies
\begin{equation}
\nabla^aF_{ab}=-4\pi J_b. \label{F_ab1}
\end{equation}
\begin{equation}
\nabla_{[a}F_{bc]}=0; \label{F_ab2}
\end{equation}

(2)From equation (\ref{F_ab2}), we know that at least there exists
1-form ${\bf A}$ locally, called electromagnetic 4-potential, so
that ${\bf F}=d{\bf A}$. The motion equation of ${\bf A}$ is
\begin{equation}
\nabla^a \nabla_aA_b-{R_b}^d A_d=-4\pi J_b; \label{A}
\end{equation}

(3)The energy-momentum tensor of electromagnetic field is
\begin{equation}
T_{ab}=\frac{1}{4\pi}(F_{ac}{F_b}^c-\frac{1}{4}g_{ab}F_{cd}F^{cd});
\label{dccT_ab}
\end{equation}

(4)The matter field tensor of the electromagnetic field is to be
studied.

(5)Let $K^a$ be the 4-wavevector of photon. By any instantaneous
observer $(P,Z^a)$ at any point $P$ of the spacetime, $K^a$ at the
point $P$ may be decomposed as
\begin{equation}
K^a=\omega Z^a+k^a; \label{Pk}
\end{equation}
where
\begin{equation}
\omega=-K^aZ_a; \label{omega}
\end{equation}
and $k^a$ are respectively the angular frequency and 3-wavevector
measured by the instantaneous observer $(P,Z^a)$.

(6)The 4-momentum $P^a$ of photon is defined as

\begin{equation}
P^a:=\hbar K^a; \label{P}
\end{equation}

(7)The world line of photon is a lightlike geodesic, and the
affine parameter $\beta$ satisfies
\begin{equation}
K^a=(\frac{\partial}{\partial \beta})^a; \label{beta}
\end{equation}

(8)By any instantaneous observer $(P,Z^a)$, $P^a$ at any point $P$
of the spacetime may be decomposed as
\begin{equation}
P^a=EZ^a+p^a. \label{Ep}
\end{equation}
where
\begin{equation}
E=\hbar \omega, \label{E}
\end{equation}
and
\begin{equation}
p^a=\hbar k^a;; \label{p}
\end{equation}
They are respectively the energy and 3-momentum of the photon.

\subsubsection{THE CONSERVED QUANTITY OF THE SCHWARZSCHILD SPACETIME}

We can select Schwarzschild coordinates so that the parameter
equations of world line $\gamma(\tau)$ of free particle
are\cite{liangchanbin}
\begin{equation}
t=t(\tau),r=r(\tau),\theta=\frac{\pi}{2},\varphi=\varphi(\tau).
\label{zyzdsjxcsbd}
\end{equation}

Let $U^a=(\frac{\partial }{\partial \tau})^a$ be the tangent
vector of a timelike geodesic $\gamma(\tau)$ and $U^a\equiv
K^a=(\frac{\partial }{\partial \beta})^a$ be the tangent vector of
a lightlike geodesic. Because $\xi^a\equiv(\frac{\partial
}{\partial t})^a$ and $\xi_{\varphi}^a\equiv(\frac{\partial
}{\partial \varphi})^a$ are the Killing vector fields of the
Schwarzschild spacetime, we can define two constants on the
timelike geodesic and the lightlike geodesic which are conserved
quantities.
\begin{equation}
E:=-g_{ab}(\frac{\partial }{\partial t})^a(\frac{\partial
}{\partial \tau})^b
=(1-\frac{2M_G}{r}\frac{1}{f(P)})\frac{dt}{d\tau}. \label{E}
\end{equation}
\begin{equation}
L:=g_{ab}(\frac{\partial }{\partial \varphi})^a(\frac{\partial
}{\partial \tau})^b =r^2\frac{d\varphi}{d\tau}. \label{L}
\end{equation}
For photon, $\tau\equiv \beta$ in the above equation.
$\xi^a\equiv(\frac{\partial }{\partial t})^a$ is the static
Killing vector field of Schwarzschild spacetime. The 4-velocity of
a static observer is
\begin{equation}
Z^a=\chi^{-1}\xi^a. \label{jtgz4s}
\end{equation}
where, $\chi\equiv (-\xi^b\xi_b)^{1/2}$. from
\begin{equation}
E=-\xi_aU^a=\frac{\chi}{f(P)m_G}E_I. \label{E1}
\end{equation}
We know that the conserved quantity $E$ of a free particle is the
energy $E_I$ per unit rest inertial mass measured by a static
observer in infinity. For free photons, from equations (\ref{P})
and (\ref{Ep}) we can obtain
\begin{equation}
E:=-g_{ab}\xi^aK^b=-g_{ab}\xi^aP^b/\hbar=\chi E_{ph}/\hbar.
\label{phE}
\end{equation}
Where, $E_{ph}=\hbar \omega$ is the energy of photon measured by a
static observer $Z^a$. From equation (\ref{phE}), we know that
$\hbar E$ of free photons is the photon energy measured by
infinity static observer.

Let $P\in \gamma(\tau)$, $Z^a$ is the static observer at point
$P$. Normalizing the coordinate basis at the point $P$ gives the
orthonormal tetrad in tangent space $V_P$ at the point
$P$\cite{liangchanbin}.
\begin{equation}
\begin{array}{ll}
(e_0)^a\equiv
(1-\frac{2M_G}{r}\frac{1}{f(P)})^{-1/2}(\frac{\partial}{\partial
t})^a,
(e_1)^a\equiv (1-\frac{2M_G}{r}\frac{1}{f(P)})^{1/2}(\frac{\partial}{\partial r})^a\\
(e_2)^a\equiv r^{-1}(\frac{\partial}{\partial \theta})^a,
(e_3)^a\equiv r^{-1}(\frac{\partial}{\partial \varphi})^a.
\end{array}
\label{4biaojia}
\end{equation}
Its dual 4-frame is
\begin{equation}
\begin{array}{ll}
(e^0)_a\equiv (1-\frac{2M_G}{r}\frac{1}{f(P)})^{1/2}(dt)_a,
(e^1)_a\equiv (1-\frac{2M_G}{r}\frac{1}{f(P)})^{-1/2}(dr)_a\\
(e^2)_a\equiv r(d\theta)_a, (e^3)_a\equiv r(d\varphi)_a.
\end{array}
\label{duiou4biaojia}
\end{equation}
The angular momentum of free particle $\gamma(\tau)$ with respect
to the static observer is defined (taking equation
(\ref{ryskdlxfc2}) or (\ref{ryskdlxfc1}) and ignoring $m_{in}$) as
\begin{equation}
j^a:={\varepsilon^a}_{bc}\gamma f(P)m_Gr^bu^c.
\label{ziyouzhidianjiaodongliang}
\end{equation}
where, $r^b\equiv r(e_1)^b$. Then we have
\begin{equation}
j_a=\varepsilon_{abc}\gamma f(P)m_Gr^bu^c =-\gamma
f(P)m_Gru^3(e^2)_a. \label{ziyouzhidianjiaodongliang1}
\end{equation}
The magnitude of the angular momentum is
\begin{equation}
j:=|\gamma f(P)m_Gru^3|=f(P)m_G|r^2\frac{d\varphi}{d\tau}|
=f(P)m_G|L|. \label{ziyouzhidianjiaodongliangdx}
\end{equation}
That is, $|L|$ is the magnitude of the 3-angular momentum with
respect to the static reference frame per unit rest inertial mass
of a free particle.The change of $f(P)$ with $P$ can be ignored in
a small space range. If $L$ is conserved, $j$ is approximately
also conserved. $\hbar L$ is the angular momentum of a photon in
lightlike geodesic\cite{wald}.

Let $U^a=(\frac{\partial }{\partial \tau})^a$ or $U^a\equiv
K^a=(\frac{\partial }{\partial \beta})^a$ is the tangent vector on
the world line $\gamma(\tau)$ of a free particle or the lightlike
geodesic. Define
\begin{equation}
-\kappa :=g_{ab}U^aU^b
=-(1-\frac{2M_G}{r}\frac{1}{f(P)})(\frac{dt}{d\tau})^2
+(1-\frac{2M_G}{r}\frac{1}{f(P)})^{-1}(\frac{dr}{d\tau})^2
+r^2(\frac{d\varphi}{d\tau})^2. \label{kappa1}
\end{equation}
Substituting equations (\ref{E}) and (\ref{L}) into equation
(\ref{kappa1}), we have
\begin{equation}
-\kappa =-(1-\frac{2M_G}{r}\frac{1}{f(P)})^{-1}E^2
+(1-\frac{2M_G}{r}\frac{1}{f(P)})^{-1}(\frac{dr}{d\tau})^2
+\frac{L^2}{r^2}. \label{kappa2}
\end{equation}
In above equation, for a photon $\equiv \tau\equiv \beta$ and
$\kappa=\left\{
\begin{array}{rl}
1,&{\hbox{the world line of a free particle}}\\
0,&{\hbox{the lightlike geodesic}}
\end{array}
\right.$

\subsubsection{GRAVITATIONAL REDSHIFT}

Let $G$ and $G'$ be two observers of arbitrary stationary
reference frame in arbitrary stationary spacetime, and the photon
emitted at the moment $P$ by $G$ reaches $G'$\cite{liangchanbin}
at the moment $P'$. Let $Z^a$ represent the 4-velocity of the
observer, and $K^a$ represent the 4-wavevector of the photon. Then
from equation (\ref{omega}), we know that the angular frequencies
of the photon with respect to the stationary observers at $P$ and
$P'$ are respectively
\begin{equation}
\omega=-(K^aZ_a)|_P,\omega'=-(K^aZ_a)|_P'. \label{omega1}
\end{equation}
The world line of the stationary observer coincides with the
integral curves of the Killing vector field ${\xi}^a$, namely
equation (\ref{jtgz4s}) applicable. As $K_a{\xi}^a$ is a constant
in the lightlike geodesic, so
\begin{equation}
\frac{\omega'}{\omega}=\frac{\chi}{\chi'}. \label{omega2}
\end{equation}
For Schwarzschild spacetime
\begin{equation}
\chi^2=-\xi^b\xi_b=-g_{00}= 1-\frac{2M_G}{r}\frac{1}{f(P)}.
\label{chi}
\end{equation}
then
\begin{equation}
\frac{\lambda'}{\lambda}=(1-\frac{2M_G}{r'}\frac{1}{f(P')})^{1/2}
(1-\frac{2M_G}{r}\frac{1}{f(P)})^{-1/2}. \label{lambda1}
\end{equation}
Here, $f(P)$ and $f(P')$ are considered to be equal.

\subsubsection{PERIHELION PRECESSION}

In Newton's theory, from equation (\ref{ndfc}) we have
\begin{equation}
{\bf F}\cdot d{\bf r}=m\frac{d{\bf V}}{dt}\cdot d{\bf r} +{\bf
V}\cdot d{\bf r}\frac{dm}{dt} =d(\frac{1}{2}mv^2)+v^2dm.
\label{gongneng}
\end{equation}
That is, the kinetic energy theorem of Newton's theory is not
established. The mechanical energy of the system, without external
forces and non-conservative internal forces, is not conserved.
\begin{equation}
{\bf r}\times {\bf F}={\bf r}\times \frac{d(m{\bf V})}{dt}
=\frac{d({\bf r}\times m{\bf V})}{dt} =\frac{d{\bf p}_L}{dt}.
\label{jdldl}
\end{equation}
in which, ${\bf p}_L\equiv {\bf r}\times m{\bf V}$ is the angular
momentum of a particle with respect to the origin. Then the
theorem of angular momentum and the law of angular momentum
conservation are still valid.

The angular momentum ${\bf p}_L={\bf r}\times m{\bf V}=f({\bf
r})m_G{\bf r}\times {\bf V}$ of the Mercury is conserved in the
solar gravitational field. Then the $p_L=f({\bf
r})m_Gr^2\frac{d\varphi}{dt}$ is constant. Thus
\begin{equation}
L\equiv r^2\frac{d\varphi}{dt} =p_L/[f({\bf r})m_G]. \label{jdl}
\end{equation}
is approximately a constant. When $\nabla f({\bf r})$ is ignored,
the mechanical energy of the Mercury is approximately conserved.
Namely
\begin{equation}
A=\frac{1}{2}m(u_r^2+u_{\varphi}^2)+(-\frac{M_Gm_G}{r}).
\label{jxnsh}
\end{equation}
$A$ is a constant. By equations (\ref{jdl}) and (\ref{jxnsh}) we
can obtain
\begin{equation}
\frac{d^2\mu}{d\varphi^2}+\mu=M_G/[f({\bf r})L^2]. \label{ndsxfc}
\end{equation}
in which, $\mu\equiv 1/r$. The solution of equation (\ref{ndsxfc})
is
\begin{equation}
\mu(\varphi)=\frac{M_G}{f({\bf r})L^2}(1+ecos\varphi).
\label{ndsxfcjie}
\end{equation}
Where, $e$ is the integral constant and is the eccentricity.

Considering the case of the curved spacetime, ignoring the
variation of $f(P)$ with respect to $P$, and taking $\kappa=1$ in
equation (\ref{kappa2}), from equations (\ref{kappa2}) and
(\ref{L}) we can obtain
\begin{equation}
\frac{d^2\mu}{d\varphi^2}+\mu=M_G/[f(P)L^2]+\frac{3M_G}{f(P)}\mu^2.
\label{wqsksxfc}
\end{equation}
where $\mu\equiv 1/r$. Equation (\ref{ndsxfcjie}) can be used as
the 0-order approximate solution of equation (\ref{wqsksxfc}),
i.e.
\begin{equation}
\mu_0(\varphi)=\frac{M_G}{f({\bf r})L^2}(1+ecos\varphi).
\label{wqsksxfcljjs}
\end{equation}
The 1-order approximation of equation (\ref{wqsksxfc}) is
\begin{equation}
\frac{d^2\mu_1}{d\varphi^2}+\mu_1=M_G/[f(P)L^2]+\frac{3M_G}{f(P)}\mu_0^2
=M_G/[f(P)L^2]+\frac{3M_G^3}{f^3(P)L^4}(1+2ecos\varphi+e^2cos^2\varphi).
\label{wqsksxfc1jijs}
\end{equation}
Then, by using the same solving method\cite{liangchanbin} with
Einstein's case perihelion precession angle in a period can be
obtained
\begin{equation}
\Delta \varphi_P\approx 6\pi M_G^2/[f^2(P)L^2]. \label{sxjrdjdj}
\end{equation}

\subsubsection{THE BENDING OF LIGHT}

Let $\kappa=0$ in equation (\ref{kappa2}). From equations
(\ref{kappa2}) and (\ref{L}) we have
\begin{equation}
(\frac{dr}{d\varphi})^2-\frac{E^2r^4}{L^2}+r^2(1-\frac{2M_G}{r}\frac{1}{f(P)})=0.
\label{wqskxgpzfc}
\end{equation}
From equation (\ref{wqskxgpzfc}) we have
\begin{equation}
\frac{d^2\mu}{d\varphi^2}+\mu=\frac{3M_G}{f(P)}\mu^2.
\label{wqskxgpzfc1}
\end{equation}
By using the same solving method\cite{liangchanbin} as the
Einstein's case, the deflection angle of distant photon through a
vacuum spherically symmetric gravitational field can be obtained
\begin{equation}
\beta\approx \frac{4M_G}{lf(P)}. \label{xgpzj}
\end{equation}
in which, $l$ is an integral constant and it is approximately the
distance from the center of spherically symmetric gravity to the
track of the distant photon when photon track is not deflected.

The external regions of quasars, galaxies, and clusters of
galaxies are nearly the cosmic background, $f(P)\approx 0.09\sim
0.17$. Therefore, the effects of gravitational lens of quasars,
galaxies, and clusters of galaxies are the $5.9\sim 11$ times as
big as those of the Einstein's theory.

In addition, if quasar is an Active Galactic Nuclei, its mass should not be
less than the mass of the galaxy, and its effect of gravitational
lens should not be weaker than that of the galaxy, rather than
only be $5.9\sim 11$ times higher than the effects of
gravitational lens of ordinary stars.

\subsection{THE INTERIOR STATE EQUATION OF STATIC SPHERICALLY SYMMETRIC STAR}

When the stellar interior matter field is regarded as a perfect
fluid, the matter field tensor is as equation (\ref{lxltwzczl}).
Ignoring $\Lambda$ and substituting equations (\ref{jtqdcxy}) and
(\ref{lxltwzczl}) into the gravitational field equation
(\ref{ylcfc}), we have the following equations which static
spherically symmetric stellar internal metric satisfies
\begin{equation}
e^{-2B}[-(1+2\alpha)A^{\prime\prime}+(1+2\alpha)A^{\prime}
B^{\prime} -(1+2\alpha){A^{\prime}}^2-(1+2\alpha)2r^{-1}A^{\prime}
+4\alpha r^{-1}B^{\prime}-2\alpha r^{-2}]+2\alpha r^{-2}
=-\tilde{\kappa}\rho. \label{jtqdcnbj1}
\end{equation}
\begin{equation}
e^{-2B}[-(1+2\alpha)A^{\prime\prime}+(1+2\alpha)A^{\prime}
B^{\prime} -(1+2\alpha){A^{\prime}}^2+(1+2\alpha)2r^{-1}B^{\prime}
-4\alpha r^{-1}A^{\prime}-2\alpha r^{-2}]+2\alpha r^{-2}
=\tilde{\kappa}f_1(p,\rho). \label{jtqdcnbj2}
\end{equation}
\begin{equation}
-e^{-2B}[(1+2\alpha)r^{-2}+(1+4\alpha)r^{-1}(A^{\prime}-B^{\prime})
+2\alpha A^{\prime\prime}-2\alpha A^{\prime} B^{\prime} +2\alpha
{A^{\prime}}^2]+(1+2\alpha)r^{-2} =\tilde{\kappa}f_1(p,\rho).
\label{jtqdcnbj3}
\end{equation}
Where, $A^{\prime}\equiv \frac{dA}{dr}$, $A^{\prime\prime}\equiv
\frac{d^2A}{dr^2}$, and $B^{\prime}\equiv \frac{dB}{dr}$. When
$\alpha=-\frac{1}{2}$,$\tilde{\kappa}=8\pi$, and $f_1(p,\rho)=p$,
the equations (\ref{jtqdcnbj1}), (\ref{jtqdcnbj2}), and
(\ref{jtqdcnbj3}) return to the case of
Einstein\cite{liangchanbin}. Static spherically symmetric stellar
internal state is decided by the four functions $A(r)$, $B(r)$,
$\rho(r)$, and $p(r)$, which satisfy the three equations
(\ref{jtqdcnbj1}), (\ref{jtqdcnbj2}), (\ref{jtqdcnbj3}), and the
state equation $F(p,\rho)=0$.

Let us discuss the following

(1)From equations (\ref{jtqdcnbj1}), (\ref{jtqdcnbj2}), and
(\ref{jtqdcnbj3}), we can obtain
\begin{equation}
2r^{-1}(B^{\prime}+A^{\prime})e^{-2B}
=\tilde{\kappa}[f_1(p,\rho)+\rho]. \label{jtqdcnbj4}
\end{equation}
\begin{equation}
e^{-2B}[-(1+2\alpha)r^{-1}B^{\prime}+(1+6\alpha)r^{-1}A^{\prime}
+(1+4\alpha)r^{-2}]-(1+4\alpha)r^{-2} =-\tilde{\kappa}f_1(p,\rho).
\label{jtqdcnbj5}
\end{equation}
From equations (\ref{jtqdcnbj4}) and (\ref{jtqdcnbj5}), we have
\begin{equation}
\frac{d}{dr}(re^{-2B})=1-\frac{3(1+2\alpha)}{2(1+4\alpha)}
\tilde{\kappa}r^2f_1(p,\rho)-\frac{1+6\alpha}{2(1+4\alpha)}
\tilde{\kappa}\rho r^2. \label{jtqdcnbj6}
\end{equation}
In the non-relativistic case, $p\ll\rho$. In the static case,
$f_2(\rho)=0$. By integrating equation (\ref{jtqdcnbj6}), we can
obtain
\begin{equation}
e^{-2B}\approx
1-\frac{1+6\alpha}{1+4\alpha}\frac{\tilde{\kappa}}{8\pi}
\frac{m(r)}{r}. \label{jtqdcnbj7}
\end{equation}
where
\begin{equation}
m(r)\equiv 4\pi f^2({\bf r})\int_0^r\frac{\rho_G(r)}{f({\bf
r})}r^2dr. \label{jtqdcnbj8}
\end{equation}
$\rho_G(r)$ is the gravitational mass density of the star.
\begin{equation}
g_{11}(r)=e^{2B}\approx
[1-\frac{1+6\alpha}{1+4\alpha}\frac{\tilde{\kappa}}{8\pi}
\frac{m(r)}{r}]^{-1}. \label{jtqdcnbj9}
\end{equation}

(2)$\nabla^aT_{ab}=0$ is approximately established in the case of
Newton approximation. Thus, we obtain
\begin{equation}
\frac{dp}{dr}\approx -(p+\rho)\frac{dA}{dr}. \label{dpdr}
\end{equation}
In the non-relativistic case, $p\ll\rho$. In the static case,
$f_2(\rho)=0$. From equations (\ref{jtqdcnbj4}),
(\ref{jtqdcnbj9}), and (\ref{dpdr}), we can obtain
\begin{equation}
\frac{dp}{dr}\approx\rho
B'-\frac{1}{2}\tilde{\kappa}\rho^2re^{-2B}. \label{jtqdcnbj10}
\end{equation}
Considering that the inertial mass density of the star is nearly
uniform, namely, $\rho\sim \overline{\rho}$, and the change of
$f({\bf r})$ with respect to $\bf r$ can be ignored, then we have
$m(r)\approx \frac{4\pi r^3}{3}\rho$. From equation
(\ref{jtqdcnbj7}) we obtain
\begin{equation}
B'\approx \frac{1+6\alpha}{1+4\alpha}\frac{\tilde{\kappa}}{16\pi}
(4\pi\rho r-\frac{m}{r^2})e^{2B}. \label{jtqdcnbj11}
\end{equation}
From equations (\ref{jtqdcnbj7}), (\ref{jtqdcnbj10}),
(\ref{jtqdcnbj11}), and taking (\ref{kappa}) into account, we can
obtain
\begin{equation}
\frac{dp}{dr}\approx -\frac{m_G\rho_G}{r^2}. \label{jtqdcnbj12}
\end{equation}
That is the results of Newtonian mechanics. The pressure $p>0$ is
the exclusion effect in the Newtonian's theory, but the positive
pressure of $p>0$ produces gravitational effects in Einstein's
gravitational field equation. Therefore, it may be unreasonable
that the Newton's form like (\ref{jtqdcnbj12}) appears in the
Einstein's theory.

(3)From equation (\ref{jtqdcnbj5}), we have
\begin{equation}
A'=\frac{1+2\alpha}{1+6\alpha}B'
-\frac{1}{1+6\alpha}\tilde{\kappa}rf_1(p,\rho)e^{2B}
+\frac{1+4\alpha}{1+6\alpha}r^{-1}(e^{2B}-1). \label{jtqdcnbj13}
\end{equation}
Ignoring the item $f_1(p,\rho)$ in equation (\ref{jtqdcnbj13}) and
taking up to 1-order approximation of $\frac{m(r)}{r}$, we can
obtain
\begin{equation}
A'\approx\frac{1}{3}\frac{1+3\alpha}{1+4\alpha}\tilde{\kappa}\rho
r. \label{A'}
\end{equation}
Considering that the inertial mass density of the star is nearly
uniform and integrating equation (\ref{A'}), we can obtain
\begin{equation}
g_{00}=-e^{2A}\approx -Ce^{\frac{1+3\alpha}{1+4\alpha}
\frac{\tilde{\kappa}}{4\pi}\frac{m(r)}{r}}. \label{g_{00}}
\end{equation}
$C$ is an integral constant. In internal vacuum region of uniform
shell, $\rho=0$, from equations (\ref{jtqdcnbj9}) and
(\ref{g_{00}}) we know that the metric of the region is Minkowski
metric.

(4)From the gravitational field equation of $\Lambda=0$
\begin{equation}
R_{ab}+\alpha Rg_{ab}=\tilde{\kappa}M_{ab}. \label{ylcfc1}
\end{equation}
and $\nabla^aG_{ab}=0$, considering the approximation of
$\nabla^a\tilde{\kappa}\approx 0$, we have
\begin{equation}
(\frac{\partial }{\partial r})^b\nabla^aM_{ab}\approx
\frac{1+2\alpha}{2\tilde{\kappa}}\frac{dR}{dr}. \label{M}
\end{equation}
From equation (\ref{lxltwzczl}) we can obtain
\begin{equation}
(\frac{\partial }{\partial r})^b\nabla^aM_{ab}=
(\rho+f_1(p,\rho))\frac{dA}{dr}+\frac{df_1(p,\rho)}{dr}.
\label{M1}
\end{equation}
From equations (\ref{f_1}), (\ref{kappa}), (\ref{jtqdcnbj1}),
(\ref{jtqdcnbj2}), (\ref{jtqdcnbj13}), (\ref{A'}), (\ref{M}),
(\ref{M1}), and considering Newton approximation and $\rho\sim
\overline{\rho}$, i.e. $\frac{d}{dr}\frac{m}{r^3}\sim
\frac{d\rho}{dr}\approx 0$, we have
\begin{equation}
\frac{6\alpha-1}{2(1+6\alpha)}\hat{\beta}\frac{dp}{dr}
 \approx -\frac{m_G\rho_G}{r^2}.
\label{M4}
\end{equation}
Comparing with equation (\ref{jtqdcnbj12}) we can obtain
\begin{equation}
\hat{\beta}=\frac{2(1+6\alpha)}{6\alpha-1}. \label{beta}
\end{equation}

(5)To avoid space-time singularity as much as possible in our
theory, it is required that there be the observer whose 4-velocity
is $Z^a$ so that $R_{ab}Z^aZ^b<0$ is satisfied, thereby the
condition of singularity theorem does not satisfy. In equation
(\ref{R_{ab}Z^aZ^b}), $\gamma\geq1$. if $R_{ab}Z^aZ^b<0$ is
established, $\gamma$ should be as small as possible. Let
$Z^a=U^a$, i.e. $\gamma=1$. Ignoring item $\Lambda$, then from
equation (\ref{R_{ab}Z^aZ^b}), we know that the establishment of
$R_{ab}Z^aZ^b<0$ means
\begin{equation}
\frac{3\alpha+1}{4\alpha+1}\tilde{\kappa}\rho
+\tilde{\kappa}\frac{3\alpha}{4\alpha+1}f_1(p,\rho)<0.
\label{jtqdcnbj14}
\end{equation}
It can be seen from equation (\ref{yzdlxfc3}) later in cosmology
that equation (\ref{jtqdcnbj14}) means the accelerating expansion
of the universe. The first mass item of equation
(\ref{jtqdcnbj14}) is the attraction effect, then
$\frac{3\alpha+1}{4\alpha+1}\tilde{\kappa}>0$. The second pressure
and inertia item is the exclusion effect, then
$\tilde{\kappa}\frac{3\alpha}{4\alpha+1}\hat{\beta}<0$. Under the
condition $\kappa<0$, we have $\frac{3\alpha+1}{4\alpha+1}<0$ and
$-\frac{1}{3}<\alpha<-\frac{1}{4}$. From the cosmology section we
know that to get the accelerating expansion evolution equation of
the universe, under the condition $\alpha>-\frac{1}{3}$, there
must be $-\frac{1}{4}<\alpha<0$. As such, there are only
$\kappa>0$ and $\frac{3\alpha+1}{4\alpha+1}>0$. Further more there
are $\alpha<-\frac{1}{3}$ or $\alpha>-\frac{1}{4}$. While
$\alpha<-\frac{1}{3}$,
$\tilde{\kappa}\frac{3\alpha}{4\alpha+1}\hat{\beta}=
\tilde{\kappa}\frac{3\alpha}{4\alpha+1}\frac{2(1+6\alpha)}{6\alpha-1}>0$,
then it is $-\frac{1}{4}<\alpha<0$. From
$\frac{3\alpha}{4\alpha+1}\hat{\beta}=
\frac{3\alpha}{4\alpha+1}\frac{2(1+6\alpha)}{6\alpha-1}<0$ we have
$\alpha<-\frac{1}{6}$. Therefore, there must be
$-\frac{1}{4}<\alpha<-\frac{1}{6}$.

(6) Due to $\frac{1+6\alpha}{1+4\alpha}<0$ in equation
(\ref{jtqdcnbj9}), from equations (\ref{jtqdcnbj9}) and
(\ref{g_{00}}) we know that there are no horizon and one-way
membrane region (i.e. the region of collapse) in the interior of
spherically symmetric star. In the evolution process of
spherically symmetric star, even if the matter of the star
contracts and gets into the event horizon of Schwarzschild vacuum
solution, the collapse singularity not necessarily will form. But
due to the energy loss of the Hawking radiation, the matter will
continue to shrink so that the divergent singularity may be
formed.If some aspects of quantum field theory are not correct so
that there is no the Hawking radiation, then the matter of the
star will eventually stop the contraction. In this case although
the spherically symmetric star has evolved as a black hole, there
is no singularity within the black hole. The reason of no
singularity in the black hole is that in some stage of the
evolution in spherically symmetric star, the second item of
equation (\ref{jtqdcnbj14}) leads to that $R_{ab}U^aU^b<0$ is
satisfied for the co-moving observer $U^a$ with the matter of the
star and the conditions of singularity theorem do not satisfy. But
please pay attention to that on the surface of star, metrics
inside and outside the star are not continuous. It seems that we
can choose $\alpha$ so that
$[1-\frac{1+6\alpha}{1+4\alpha}\frac{\tilde{\kappa}}{8\pi}
\frac{m(r)}{r}]^{-1}$ of equation (\ref{jtqdcnbj9}) is equal to
the $ [1- 2\frac{M_G}{rf(r)}]^{-1}$ of Schwarschild vacuum
solutions at $r=R$. Namely,
$\frac{1+6\alpha}{1+4\alpha}\frac{\tilde{\kappa}}{8\pi}
\frac{m(R)}{R}=\frac{2M_G}{Rf(R)}$. Wherein $M_G\equiv
4\pi\int_0^R\rho_G(r)[1-\frac{1+6\alpha}{1+4\alpha}\frac{\tilde{\kappa}}{8\pi}
\frac{m(r)}{r}]^{-\frac{1}{2}}r^2dr$ is the gravitational mass of
the star. But $\alpha$, thus obtained, is dependent on $\rho_G(r)$
and $f(r)$. Therefore, it is inevitable that, on the star surface,
the metrics inside and outside are not continuous. If vacuum is
regarded as a special case of $\rho_G(r)=0$, and the upper limit
$r$ of integrations of $m(r)$ in equations (\ref{jtqdcnbj9}) and
(\ref{g_{00}}) enters the vacuum region, then the metric thus
obtained will be unlike the vacuum metric. This shows that matter
region and the vacuum region should be described by using the
corresponding field equations respectively and the field equations
can not generally cross the interface of the matter and the
vacuum. The metrics on the interface of matter and vacuum are not
necessarily discontinuous. For example, the metrics on the
interface of the vacuum inside a uniform matter spherical shell
and the spherical shell matter are continuous. It is associated
with the pressure from the physical point of view that the metrics
on the interface of matter and vacuum may not be continuous.

\section{THE EVOLUTION OF THE UNIVERSE}
\subsection{THE DYNAMIC EQUATION OF THE UNIVERSE}
\subsubsection{THE ESSENTIAL EQUATIONS OF EVOLUTION OF $a(t)$}

Let us assume that the cosmological principle is applicable, then
the metric of the universe is the Robertson-Walker metric
(\ref{R-W})\cite{liangchanbin}. The contents of universe can be
regarded as a perfect fluid. Substituting Robertson-Walker metric
(\ref{R-W}) and the matter field tensor (\ref{lxltwzczl}) of
perfect fluid into the gravitational field equation (\ref{ylcfc}),
we can obtain
\begin{equation}
-\frac{3\ddot{a}}{a}=\tilde{\kappa}\frac{(1+3\alpha)\rho +3\alpha
f_1(p,\rho)}{1+4\alpha}+\frac{\Lambda}{1+4\alpha}.
\label{yzdlxfc1}
\end{equation}
\begin{equation}
a\ddot{a}+2\dot{a}^2+2k=\tilde{\kappa}\frac{\alpha\rho
+(1+\alpha)f_1(p,\rho)}{1+4\alpha}a^2-\frac{\Lambda}{1+4\alpha}a^2.
\label{yzdlxfc2}
\end{equation}
From equation (\ref{yzdlxfc1}) we know that if let $\alpha
>-\frac{1}{4}$, cosmological constant $\Lambda$ has an attraction rather
than repulsion effect. Considering the case of cosmological
constant $\Lambda=0$, the equations (\ref{yzdlxfc1}) and
(\ref{yzdlxfc2}) become
\begin{equation}
-\frac{3\ddot{a}}{a}=\tilde{\kappa}\frac{(1+3\alpha)\rho +3\alpha
f_1(p,\rho)}{1+4\alpha}. \label{yzdlxfc3}
\end{equation}
\begin{equation}
a\ddot{a}+2\dot{a}^2+2k=\tilde{\kappa}\frac{\alpha\rho
+(1+\alpha)f_1(p,\rho)}{1+4\alpha}a^2. \label{yzdlxfc4}
\end{equation}
Equations (\ref{yzdlxfc3}) and (\ref{yzdlxfc4}) are the essential
equations to determine the evolution of the scale factor $a(t)$ of
the universe. When $\alpha=-\frac{1}{2}$, $\tilde{\kappa}=8\pi$,
and $f_1(p,\rho)=p$, the equations (\ref{yzdlxfc3}) and
(\ref{yzdlxfc4}) return to the Einstein's
theory\cite{liangchanbin}.

\subsubsection{THE CRITICAL DENSITY OF INERTIAL MASS OF THE UNIVERSE}

Discussing three cases below

(1)$\ddot{a}>0$

Due to $-\frac{1}{4}<\alpha<0$, and from equation
(\ref{yzdlxfc3}), we have
\begin{equation}
(1+3\alpha)\rho+3\alpha f_1(p,\rho)<0. \label{ljzl1}
\end{equation}
From equations (\ref{yzdlxfc3}), (\ref{yzdlxfc4}), and
(\ref{ljzl1}), we can obtain
\begin{equation}
\rho<\rho_c+\frac{-6\alpha k}{\tilde{\kappa}Ga^2}. \label{ljzl6}
\end{equation}
Where, $\rho_c\equiv \frac{-16\pi\alpha f^2(P)}{\kappa}\rho_{cE}$
is the critical density of cosmic inertial mass. $\rho_{cE}\equiv
\frac{3H^2}{8\pi G}$ is the critical density of the universe mass
in Einstein's theory.

(2)$\ddot{a}=0$
\begin{equation}
\rho=\rho_c+\frac{-6\alpha k}{\tilde{\kappa}Ga^2}. \label{ljzl7}
\end{equation}

(3)$\ddot{a}<0$
\begin{equation}
\rho>\rho_c+\frac{-6\alpha k}{\tilde{\kappa}Ga^2}. \label{ljzl8}
\end{equation}
Let us take $k=0$, namely consider the flat universe. $\ddot{a}>0$
results in $\rho<\rho_c$; $\ddot{a}=0$ results in $\rho=\rho_c$;
$\ddot{a}<0$ results in $\rho>\rho_c$. That is, when the universe
is flat, $\rho$ takes its value in an interval containing
$\rho_c$; there is no the flatness problem in Einstein's theory.
If we take $\alpha=-\frac{1}{5}$, today $f(P)=\alpha_U=0.09\sim
0.1$, then the critical density of the universe inertial mass
today is $\rho_c\sim 0.018\rho_{cE}$ which is approximately
$1.8\%$ of the value in Einstein's theory. Today the universe is
accelerating expansion, hence, the inertial mass density of the
universe is $\rho<1.8\%\rho_{cE}$.

\subsection{THE EVOLUTION OF $a(t)$}

Substituting
\begin{equation}
f^2(P)=f(P)\frac{4\pi K\rho_G}{\delta^2}=\frac{4\pi
K}{\delta^2}\rho. \label{f(P)}
\end{equation}
and $\tilde{\kappa}=\kappa/\frac{4\pi K}{\delta^2}\rho$ into
equations (\ref{yzdlxfc3}) and (\ref{yzdlxfc4}), we have
\begin{equation}
\frac{1+2\alpha}{\alpha}\frac{1+3\alpha}{1+4\alpha}
\frac{\ddot{a}}{\beta a}
+\frac{2(1+3\alpha)}{1+4\alpha}\frac{1}{\beta}(\frac{\dot{a}}{a})^2
+\frac{2(1+3\alpha)}{1+4\alpha}\frac{k}{\beta
a^2}+\frac{1}{3\alpha}=0. \label{ayanhuafc2}
\end{equation}
where $\beta \equiv\frac{\delta^2}{K}$. Let us take into account
the case of $k=0$. Due to $\alpha>-\frac{1}{3}$ and solving
equation (\ref{ayanhuafc2}), we can obtain
\begin{equation}
a(t)=A|\cos[\omega(t-t_0)+\varphi]|^n. \label{aj}
\end{equation}
Where,
$A=a_0/|\cos[\arctan\sqrt{\frac{3(1+3\alpha)}{\beta}}y_0]|^n$,
$t_0$ is the value of $t$ at a certain moment, $y_0\equiv
\frac{\dot{a}}{a}|_{t=t_0}$,$a_0\equiv a(t_0)$, $\omega\equiv
\frac{1+4\alpha}{1+2\alpha}\sqrt{\frac{\beta}{3(1+3\alpha)}}$,
$\varphi\equiv -\arctan\sqrt{\frac{3(1+3\alpha)}{\beta}}y_0$, and
$n\equiv \frac{1+2\alpha}{1+4\alpha}$.

Without loss of generality, we consider
$-\frac{\pi}{2}\leq\omega(t-t_0)+\varphi\leq \frac{\pi}{2}$, then
we have
\begin{equation}
a(t)=A\cos^n[\omega(t-t_0)+\varphi]. \label{aj1}
\end{equation}
If $a$ experienced $a=0$ and let $a|_{t=0}=0$, then $n>0$,
$-\omega t_0+\varphi=-\frac{\pi}{2}$, and $0\leq\omega t\leq \pi$.
\begin{equation}
a(t)=A\cos^n(\omega t-\frac{\pi}{2})=A\sin^n(\omega t).
\label{aj2}
\end{equation}
\begin{equation}
\dot{a}(t)=nA\omega \sin^{n-1}(\omega t)\cos(\omega t).
\label{aj3}
\end{equation}
Hubble parameter is
\begin{equation}
H\equiv \frac{\dot{a}(t)}{a(t)}=n\omega \cot(\omega t).
\label{aj4}
\end{equation}
\begin{equation}
\ddot{a}(t)=nA\omega^2[(n-1)\sin^{n-2}(\omega t)\cos^2(\omega t)
-\sin^n(\omega t)]. \label{aj5}
\end{equation}

讨论如下几个方面 Let us discuss the following aspects

1)If we take $a|_{t=0}=0$, we can not take $t_0=0$. Otherwise the
initial condition
$y_0=\frac{\dot{a}}{a}|_{t=t_0}=H|_{t=t_0}=H|_{t=0}\to \infty$
makes no sense.

2)The stage of $\ddot{a}(t)>0$ satisfies
\begin{equation}
0<\tan^2(\omega t)<n-1. \label{jspztj}
\end{equation}
Thus $n>1$ and $-\frac{1}{4}<\alpha<0$ also can be obtained

3)The evolution of $a(t)$ is periodic, and the period is
$T=\frac{1}{2}\times 2\pi \frac{1+2\alpha}{1+4\alpha}
\sqrt{\frac{3(1+3\alpha)}{\beta}}$. In SI system, from $\beta
\equiv\frac{\delta^2}{K}$ and equations (\ref{K}) and
(\ref{yzylzlmd}) we can obtain.
\begin{equation}
T=\pi \frac{\alpha_U(1+2\alpha)}{1+4\alpha}
\sqrt{\frac{3(1+3\alpha)}{4\pi G\rho_0}}. \label{T}
\end{equation}
Wherein, $\rho_0$ is today's inertial mass density of the
universe. Let us take $\rho_0=0.5\%\rho_{cE0}$. $\rho_{cE0}\equiv
\frac{3H_0^2}{8\pi G}$ is today's critical density of the universe
mass in Einstein's theory. $H_0$ is today's Hubble parameter.
Therefore
\begin{equation}
T=\frac{20\pi \alpha_U}{H_0}n \sqrt{1+3\alpha}. \label{T1}
\end{equation}

4)Let $t_{td}$ be today's value of $t$. $t_{td}$ may be called the
age of the universe.
\begin{equation}
H_0=n\omega \cot(\omega t_{td}) =\frac{H_0}{20\alpha_U}
\frac{1}{\sqrt{1+3\alpha}} \cot[\frac{H_0}{20n\alpha_U}
\frac{t_{td}}{\sqrt{1+3\alpha}}]. \label{t_{td}}
\end{equation}
If $t_{td}$ has been measured, $\alpha$ can be obtained from
equation (\ref{t_{td}}).

5)For the convenience of discussion, let $\alpha=-\frac{1}{5}$,
then $n=3$ in the following
\begin{equation}
a(t)=A\sin^3(\omega t). \label{aj6}
\end{equation}
Let us take present $\alpha_U=0.09\sim 0.1$ and
$H_0=71km.s^{-1}.Mpc^{-1}$, then
\begin{equation}
T\approx \frac{1}{2}\times 3284.3\hbox{hundred million
years}=1642.15\hbox{hundred million years}. \label{T2}
\end{equation}
From equation (\ref{jspztj}), the conditions, which the universe
of accelerating expansion satisfy, can be obtained
\begin{equation}
0<\tan(\omega t)<\sqrt{2}. \label{jspztj1}
\end{equation}
From equation (\ref{t_{td}}) we can obtain
\begin{equation}
0<\tan(\omega t_{td})\approx \frac{1}{2\sqrt{0.4}}<\sqrt{2}.
\label{jspztj2}
\end{equation}
That is, today's universe is in a stage of accelerating expansion.
From equation (\ref{aj5}), we can also obtain
$\frac{d}{dt}\ddot{a}(t)>0$. That means today's universe expanding
acceleration is increasing. From equation (\ref{t_{td}}), the age
of the universe can be obtained
\begin{equation}
t_{td}\approx 3.49674971\times 10^{10}\hbox{years}\approx
350\hbox{hundred million years}. \label{t_{td}1}
\end{equation}

6)Because the evolution of $a(t)$ is periodic, there is no
beginning of time. Therefore, there is no horizon problem.

7)From equations (\ref{aj3}) and (\ref{aj5}), we know that
$\dot{a}(0)=0$ and $\ddot{a}(0)=0$. There is no big bang, neither
is there big crunch. The Robertson-Walker metric (\ref{R-W}) will
degenerate into timelike 1-dimensional. As such, all geodesics are
tangent to the geodesic of $t$ coordinate curve at $a(t)=0$ in the
way that $\theta$ and $\varphi$ are constants, and
$\frac{dr}{d\beta}=0$,$\frac{d^2r}{d\beta^2}=0$, and
$\frac{d^2t}{d\beta^2}=0$ ($\beta$ is the affine parameter of the
geodesics). All geodesics are complete. That is, in the universe
spacetime, there is no geodesically incomplete spacetime
singularity. But still there is the divergence singularity at
$a(t)=0$. For example, $\Re\equiv R^{\mu\nu}R_{\mu\nu}
=9(\frac{\ddot{a}}{a})^2+3a^{-4}(a\ddot{a}+2\dot{a}^2+2k)^2$,
considering the case of $k=0$, from equations (\ref{aj4}),
(\ref{aj5}), and (\ref{aj6}) we know that when $t\to 0$, $\Re\sim
t^{-4}$ is divergent. I can hardly believe that the theory of
quantum gravity could save the geodesically incomplete singularity
that is formed by the collapse or the big crunch and at which the
big bang will happen. I believe that the future quantum gravity
theory will definitely make $a(t)$ have minimum quantum $a_q$; and
case $a(t)=0$ will not happen, such that divergent singularity can
be avoided.

8)Whether the universe is radiation dominated or matter dominated,
the evolution of $a(t)$ is equation (\ref{aj2}).

\subsection{THE EVOLUTION OF $\rho$}

Let $\Delta m$ be the inertial mass in the co-moving volume
element $\Delta V$ of the cosmic perfect fluid, then
\begin{equation}
\frac{1}{\Delta m}\frac{d\Delta m}{dt}=
\frac{1}{{a^3(t)r^2\sin\theta\Delta r\Delta \theta\Delta
\varphi\rho}} \frac{d[a^3(t)r^2\sin\theta\Delta r\Delta
\theta\Delta \varphi\rho]}{dt} =\frac{1}{{a^3(t)\rho}}
\frac{d[a^3(t)\rho]}{dt}. \label{Delta m}
\end{equation}
Substituting equation (\ref{f(P)}) into (\ref{yzdlxfc3}), we can
obtain
\begin{equation}
\frac{2\ddot{a}}{\beta a}+\frac{2}{3}=\frac{1}{\rho}f_1(p,\rho).
\label{rhoyhfc1}
\end{equation}
Referring to equation (\ref{guanxinpaichili}) and taking equation
(\ref{Delta m}) into account, we can assume, in equation
(\ref{f_1}),
\begin{equation}
f_2(\rho)=-\frac{\tau}{a^3}\frac{d(a^3\rho)}{dt}\frac{\dot{a}}{a}.
\label{f_2}
\end{equation}
Where $\tau$ is a constant to be determined.

\subsubsection{THE PERIOD OF RADIATION DOMINATED}

In the radiation dominated period, $p=\frac{1}{3}\rho$,
substituting equations (\ref{beta}) and (\ref{f_2}) into
(\ref{f_1}), we can obtain
\begin{equation}
\frac{1}{\rho}f_1(p,\rho)=\frac{1}{\rho}[\hat{\beta} p+f_2(\rho)]
=\frac{2}{33}-\frac{\tau}{a^3\rho}\frac{d(a^3\rho)}{dt}\frac{\dot{a}}{a}.
\label{rhoyhfc2}
\end{equation}
Substituting equations (\ref{aj6}) and (\ref{rhoyhfc2}) into
(\ref{rhoyhfc1}), we have
\begin{equation}
\frac{3\hat{\tau}}{a^3\rho}\frac{d(a^3\rho)}{dt} =-12\omega
\cot(\omega t)-\frac{6}{11}\omega\tan(\omega t). \label{rhoyhfc3}
\end{equation}
Where $\hat{\tau}\equiv \beta\tau$. Solving equation
(\ref{rhoyhfc3}) we can obtain
\begin{equation}
\rho (t)=\rho_0 (\frac{a_0}{a})^3|\frac{\sin(\omega t_0)}
{\sin(\omega t)}|^{\frac{4}{\hat{\tau}}} |\frac{\cos(\omega t)
}{\cos(\omega t_0)}|^{\frac{2}{11\hat{\tau}}} =\rho_0
|\frac{\sin(\omega t_0)}{\sin(\omega
t)}|^{\frac{4+9\hat{\tau}}{\hat{\tau}}} |\frac{\cos(\omega t)
}{\cos(\omega t_0)}|^{\frac{2}{11\hat{\tau}}}. \label{rhoyhfc4}
\end{equation}
Where $\rho_0\equiv \rho(t_0)$.

\subsubsection{THE PERIOD OF MATTER DOMINATED}

In the matter dominated period, the pressure is $p\ll \rho$. It is
negligible.
\begin{equation}
\frac{1}{\rho}f_1(p,\rho)\approx
-\frac{\tau}{a^3\rho}\frac{d(a^3\rho)}{dt}\frac{\dot{a}}{a}.
\label{wrhoyhfc2}
\end{equation}
Substituting equations (\ref{aj6}) and (\ref{wrhoyhfc2}) into
(\ref{rhoyhfc1}), we have
\begin{equation}
\frac{3\hat{\tau}}{a^3\rho}\frac{d(a^3\rho)}{dt} =-12\omega
\cot(\omega t)-\frac{6}{5}\omega\tan(\omega t). \label{wrhoyhfc3}
\end{equation}
Solving equation (\ref{wrhoyhfc3}) we can obtain
\begin{equation}
\rho (t)=\rho_0 (\frac{a_0}{a})^3|\frac{\sin(\omega t_0)}
{\sin(\omega t)}|^{\frac{4}{\hat{\tau}}} |\frac{\cos(\omega t)
}{\cos(\omega t_0)}|^{\frac{2}{5\hat{\tau}}} =\rho_0
|\frac{\sin(\omega t_0)}{\sin(\omega
t)}|^{\frac{4+9\hat{\tau}}{\hat{\tau}}} |\frac{\cos(\omega t)
}{\cos(\omega t_0)}|^{\frac{2}{5\hat{\tau}}}. \label{wrhoyhfc4}
\end{equation}

\subsubsection{THE THERMAL HISTORY OF THE UNIVERSE}

Radiation satisfies the law of blackbody radiation
\begin{equation}
\rho =\sigma T^4. \label{fsdl}
\end{equation}
Where, $T$ is radiation temperature. $\sigma \equiv
(\pi^2/30)N_{eff}$. $N_{eff}$ is a constant coefficients
determined by the number of particle species that the rest energy
of the particles is much smaller than $k_BT$. The evolution of the
radiation temperature in the radiation dominated period can be
obtained from equations (\ref{rhoyhfc4}) and (\ref{fsdl}).
\begin{equation}
T^4(t)=\frac{\rho_0 }{\sigma } |\frac{\sin(\omega
t_0)}{\sin(\omega t)}|^{\frac{4+9\hat{\tau}}{\hat{\tau}}}
|\frac{\cos(\omega t) }{\cos(\omega
t_0)}|^{\frac{2}{11\hat{\tau}}}. \label{fswd}
\end{equation}
After the decoupling of Neutrino at the temperature $T_{\nu}$, the
ratio of neutron number $n_n$ and proton number $n_p$ is frozen at
\begin{equation}
n_n/n_p=e^{-\Delta m/T_{\nu}}. \label{n_n/n_p}
\end{equation}
Where $\Delta m\equiv m_n-m_p$为 is the difference between the
rest inertial masses of neutron and proton. The light split of
deuterium is invalid at the temperature $T_D$. Let $N_{\nu}=3$ be
the number of types of neutrino and $Y_4=0.221\sim 0.243$ be the
abundance of helium, then selecting the suitable value of the
ratio $\eta$ of nucleon number and photon number in the radiating
gas, we can determine $T_{\nu}$ and $T_D$. Let $t_{\nu}$, $t_D$,
and $t_{ph}$ denote the time of the decoupling of Neutrino, the
failure time of deuterium light split, and the time of the
decoupling of photon respectively. Let the temperature of the
decoupling of photon be $T_{ph}=3000K$ and the temperature of
today's photon background be $T_{td}=2.735K$, we can obtain the
following equations
\begin{equation}
T^4_{\nu}=\frac{\rho_0 }{\sigma } |\frac{\sin(\omega
t_0)}{\sin(\omega t_{\nu})}|^{\frac{4+9\hat{\tau}}{\hat{\tau}}}
|\frac{\cos(\omega t_{\nu})}{\cos(\omega
t_0)}|^{\frac{2}{11\hat{\tau}}}. \label{T_{nu}}
\end{equation}
\begin{equation}
T^4_D=\frac{\rho_0 }{\sigma } |\frac{\sin(\omega t_0)}{\sin(\omega
t_D)}|^{\frac{4+9\hat{\tau}}{\hat{\tau}}} |\frac{\cos(\omega t_D)
}{\cos(\omega t_0)}|^{\frac{2}{11\hat{\tau}}}. \label{T_D}
\end{equation}
\begin{equation}
T^4_{ph}=\frac{\rho_0 }{\sigma } |\frac{\sin(\omega
t_0)}{\sin(\omega t_{ph})}|^{\frac{4+9\hat{\tau}}{\hat{\tau}}}
|\frac{\cos(\omega t_{ph}) }{\cos(\omega
t_0)}|^{\frac{2}{11\hat{\tau}}}. \label{T_{ph}}
\end{equation}
\begin{equation}
T^4_{td}=\frac{\rho_0 }{\sigma } |\frac{\sin(\omega
t_0)}{\sin(\omega t_{td})}|^{\frac{4+9\hat{\tau}}{\hat{\tau}}}
|\frac{\cos(\omega t_{td}) }{\cos(\omega
t_0)}|^{\frac{2}{11\hat{\tau}}}. \label{T_{td}}
\end{equation}
Wherein, for simplicity, the temperature of today's photon
background is obtained by using the temperature equation
(\ref{fswd}) of the radiation dominated period. The inertial mass
density $\rho_{td}$ of today's universe is regarded as a known
quantity - we have the following equation
\begin{equation}
\rho_{td} =\rho_0 |\frac{\sin(\omega t_0)}{\sin(\omega
t_{td})}|^{\frac{4+9\hat{\tau}}{\hat{\tau}}} |\frac{\cos(\omega
t_{td}) }{\cos(\omega t_0)}|^{\frac{2}{5\hat{\tau}}}.
\label{rho_{td}}
\end{equation}
The 5 undetermined quantities $t_{\nu}$, $t_D$, $t_{ph}$,
$\rho_0$, and $\hat{\tau}$ can be obtained by using the five
equations (\ref{T_{nu}})$\sim$(\ref{rho_{td}}).

As there is no big bang, the expansion of the universe is very
slow in a long period of time starting from $t=0$. There is enough
time to evolve and to give the total contents of the universe
evolution by the Big Bang theory. Merely the process of the
universe evolution is not the same as the Big Bang theory. As a
rough estimate, let us assume that the time from the start of the
nucleosynthesis to today equal approximately the time of the Big
Bang theory. Then the time before nucleosynthesis is approximately
the same as the time $t_{\gamma d}$ of the neutrino decoupling.
The time is approximately $150\sim 200$ hundred million years. In
such a long time, the standard model of particle physics maybe can
give the asymmetry numbers of baryons and anti-baryons required by
the formation of today's universe.

If the high-redshift of the absorption cloud in front of a quasar
is the cosmological redshift, the absorption cloud is very
ancient. The abundance of deuterium in the absorption cloud should
be the result of primordial nucleosynthesis. The abundance of
deuterium in the absorption cloud in front of all high redshift
quasars should be essentially the same. Larger differences of the
abundance of deuterium in different absorption cloud given by
observations perhaps suggest that at least a portion of the high
redshift of absorption clouds in front of quasars may not be
cosmological redshift. But both the high redshift of absorption
clouds and the redshift of quasars may be the result of the
inertial effects as the absorption clouds and the quasars are away
from the Milky Way.

\subsubsection{THE FORMATION OF THE STRUCTURE OF THE UNIVERSE}

The expansion of the universe is very slow in a long period of
time starting from $t=0$. The duration of the radiation dominated
period is very long. It can be seen from equation (\ref{fswd})
that the radiation temperature is $T\sim |\sin(\omega
t)|^{-\frac{4+9\hat{\tau}}{4\hat{\tau}}}$. If the index
$\frac{4}{4\hat{\tau}}$ is ignored, it is $T\sim |\sin(\omega
t)|^{-\frac{9}{4}}$. For example, if in the radiation dominated
period $\sin^{\frac{9}{4}}(\omega t)<10^{-6}$, then the duration
$t\sim 10^{-\frac{24}{9}}\frac{1}{\omega }$ of the radiation
dominated period is about hundreds of millions of years. During
either the radiation dominated period or the matter dominated
period, it has enough time for the self-gravitational instability
to produce respectively large-scale or small-scale mass density
perturbations. The structure of the universe is forming in the
manner of the top-down and bottom-up respectively.

\section{THE PROBLEMS THAT MAY EXIST }

At least there are two following problems for the inertial effect.

\subsection{ABOUT THE INERTIAL MASS FORMULA}

If the inertial mass formula (\ref{gxzlgs2}) is correct, then,
compared to the hydrogen spectrum on the Earth, blue shift will be
occurred on the solar hydrogen spectrum and the pendulum period
effect shown in equation (\ref{danbaizhouqixiangduicha}) is easily
measured; but that is not the case. Therefore, there may exist
problems in the inertial mass formula (\ref{gxzlgs2}). It is
likely that equation (\ref{gxzlgs2}) is asymptotic formula only
when $r\to \infty$. In a general case, the inertial mass formula
may have the following form
\begin{equation}
M_I\equiv M_{I12}=M_{I21}=\tilde{K}
(\frac{r_0}{r})^{n(r)}M_{G1}M_{G2}exp(-\delta r). \label{ybgxzlgs}
\end{equation}
Wherein, $\tilde{K}$ and $r_0$ are constants which satisfy
$\tilde{K}r_0=K$. $n(r)$ is a dimensionless quantity which is
dependent on $r$ in some way, $0\leq n(r)\leq 1$. The smaller the
$r$ is, the closer to 0 the $n(r)$ is. Below the scale of the
solar system, it is $n(r)\approx 0$. The greater the $r$ is, the
closer to 1 the $n(r)$ is. Above the scale of the galactic scales,
it is $n(r)\approx 1$. $r_0$ and $n(r)$ should be given through
observation. When $n(r)=0$, equation (\ref{ybgxzlgs}) becomes
\begin{equation}
M_I\equiv M_{I12}=M_{I21}=\tilde{K} M_{G1}M_{G2}exp(-\delta
r)\approx \tilde{K}M_{G1}M_{G2}. \label{ybgxzlgs1}
\end{equation}
When $r=r_0$, equation (\ref{ybgxzlgs}) becomes
\begin{equation}
M_I\equiv M_{I12}=M_{I21}=\tilde{K} M_{G1}M_{G2}exp(-\delta
r_0)\approx \tilde{K} M_{G1}M_{G2}. \label{ybgxzlgs2}
\end{equation}
Therefore, $r_0$ flags the region size of $n(r)\approx 0$. In the
scale of $r_0$, $M_I$ is approximately independent to $r$. Then,
we can let $r_0$ be the scale of solar system.

When the inertial mass formula (\ref{ybgxzlgs}) replaces
(\ref{gxzlgs2}), except for the pendulum effect of equation
(\ref{danbaizhouqixiangduicha}) and the discussions about the
additional acceleration effects of the spacecraft near the Earth
and the solar system in $\S 3.1$, the discussions of the rest of
this article are applicable. If in the scales of solar system
$n(r)$ in equation (\ref{ybgxzlgs}) is not strictly equal to $0$,
then the effects of pendulum period will still exist; but they are
much weaker than the effect in equation
(\ref{danbaizhouqixiangduicha}). It can be said that the result in
equation (\ref{danbaizhouqixiangduicha}) is the upper limit of the
effect of the pendulum period, i.e., $\frac{T_P-T_A}{T_A}\leq
10^{-4}$. As long as there is a difference between the pendulum
period at perihelion and aphelion, no matter how small the
difference is, it should be due to the inertial effect.

\subsection{ABOUT THE "RETARDED" EFFECT}

If, in vacuum, the inertial interaction propagates in the vacuum
light speed $c$, there exists the "retarded" effect. The inertial
mass formula (\ref{gxzlgs2}) or (\ref{ybgxzlgs}) is only suitable
for the static case. Because the universe is expanding, the cosmic
gravitational mass density decreases with the expansion of the
universe. When we calculate the contribution to the inertial mass
of the testing particle by the cosmic background at time $t$, the
further distant the substance is, the earlier than time $t$ the
contribution to the inertial mass should be. This is the
"retarded" effect.

The equation (\ref{yzbjgxzyzl}) is only applicable to the static
universe. That is, the universe neither expands nor contracts. The
slower the universe expansion is, the smaller the "retarded"
effect is. Taking the attenuation factor $exp(-\delta r)$ into
account in equations (\ref{gxzlgs2}) and (\ref{ybgxzlgs}), the
further distant, i.e. the more "retarded", a substance is, the
smaller the contribution to the inertial mass of the testing
particle is. Therefore, the slower universe expansion is, the more
reliable the result of equation (\ref{yzbjgxzyzl}) is. Because the
gravitational mass in a volume element $\varepsilon$ does not
change with respect to $t$, i.e., $\frac{d}{dt}(\varepsilon
\rho_G)=0$, then $\dot{\rho_G}/\rho_G \sim
\dot{\varepsilon}/\varepsilon=3H$ can be obtained. If we think
that the impact on the inertial mass of the testing particle by
the substance in the distance $>1/\delta$ can be ignored, then the
relative change of the gravitational mass density in the
propagation time $\frac{1/\delta}{c}$ of the inertial interaction
in the distance $1/\delta$ is $(\dot{\rho_G}/\rho_G)
\frac{1/\delta}{c} \sim \frac{3H}{c\delta}$. If
$\frac{3H}{c\delta}\ll 1$, then we can believe that the expansion
of the universe is very slow or the "retarded" effect can be
ignored. Let $H=71km.s^{-1}.Mpc^{-1}$ today and $1/\delta\sim
\sigma r_S\sim 10^{23}m$, then $\frac{3H}{c\delta}\sim 10^{-3}$
today. The "retarded" effect can be ignored.

Furthermore, $\alpha_U$ inferred by quasars is the actual value
including the "retarded" effect. Due to the cosmological
principle, $\alpha_U$ is just a function of time, independent of
the spatial position. Multiplying equation (\ref{yzbjgxzyzl}) by
the correction factor $\gamma (t)$ due to the "retarded" effect,
we can obtain
\begin{equation}
\gamma (t)K\frac{4\pi \rho_G}{\alpha_U\delta^2}=1.
\label{xishuxiuzheng}
\end{equation}
After the universe gravitational mass density $\rho_G (t)$ is
measured, $\gamma (t)$ can be determined. If the expansion of the
universe is slow, $\gamma (t)$ is a slowly varying function of the
time $t$.

The "retarded" effect is to be further studied.

\section{ACKNOWLEDGEMENTS}

I really appreciate Professor Liu Liao, Professor Liang Canbin,
and Professor Zhao Zheng of Beijing Normal University for their
cultivation and education to me! I would like to thank my close
friend Mr. You Rong and his sweet daughter You Jing-Ya for their
help when I translate this paper into English!

\end{document}